\documentclass{article}
\usepackage[T1]{fontenc}
\usepackage[latin1]{inputenc}

\makeatletter

\newcommand{\LyX}{L\kern-.1667em\lower.25em\hbox{Y}\kern-.125emX\spacefactor1000}

\newcommand{\lyxaddress}[1]{
  \par {\raggedright #1 
  \vspace{1.4em}
  \noindent\par}
}

\makeatother

\begin{document}

\title{Non-Fermi Liquid in a Truncated Two-Dimensional Fermi Surface}

\author{A.Ferraz\protect\( ^{*}\protect \)}

\maketitle

\lyxaddress{Center for Theoretical Studies-Theoretische Physik; ETH-Hongerberg; CH-8093
Zurich and Institut de Physique Theorique, Universite de Fribourg, Perolles,
Chemin du Musee 3, CH-1700 Fribourg, Switzerland}

\begin{abstract}
Using perturbation theory and the field theoretical renormalization group approach
we consider a two-dimensional anisotropic truncated Fermi Surface\( \left( FS\right)  \)
with both flat and curved sectors which approximately simulates the ``cold''
and ``hot'' spots in the cuprate superconductors. We calculate the one-particle
two-loop irreducible functions \( \Gamma ^{\left( 2\right) } \) and \( \Gamma ^{\left( 4\right) } \)
as well as the spin, the charge and pairing response functions up to one-loop
order. We find non-trivial infrared stable fixed points and we show that there
are important effects produced by the mixing of the existing scattering channels
in higher order of perturbation theory. Our results indicate that the ``cold
`` spots are turned into a non-Fermi liquid with divergents \( \partial \Sigma _{0}/\partial p_{0} \)
and \( \partial \Sigma _{0}/\partial \overline{p} \), a vanishing \( Z \)
and either a finite or zero ``Fermi velocity'' at \( FS \) when the effects
produced by the flat portions are taken into account.
\end{abstract}

\section{Introduction}

The appearance of high-T\( _{c} \) superconductivity focused everyone's attention
on the properties of strongly interacting two-dimensional electron systems.
Basically the high-T\( _{c} \) cuprates are characterized by a doping parameter
which regulates the amount of charge concentration in the CuO\( _{2} \) planes.
As one varies the doping concentration and temperature one finds an antiferromagnetic
phase, a pseudo-gap phase, an anomalous metallic phase and a d-wave superconductor.

The standard model to describe these phenomena is the the two-dimensional (\( 2d \))
Hubbard model. Starting either from the so-called weak coupling limit or from
the large \( U \) limit instead one can reproduce at least in qualitative terms
all these phases by varying only a small number of appropriate parameters\cite{Schrieffer}.
In particular, for the underdoped and optimally doped compounds, motivated by
the experimental results coming from angle-resolved photoemission experiments
\( \left( ARPES\right)  \) which demonstrated among other things the presence
of an anisotropic electronic spectra characterized by a pseudogap and flat bands
in \( \mathbf{k} \)-space several workers have related some of these anomalies
to the existence of a non-conventional Fermi Surface \( \left( FS\right)  \)
in these materials\cite{Rice}. As is well known for the half-filled \( 2d \)-
Hubbard model the \( FS \) being perfectly square the perfect nesting and the
presence of van Hove singular points allow the mapping of this system onto perpendicular
sets of one-dimensional chains\cite{Luther} producing infrared divergences
in both particle-particle and particle-hole channels, already at one-loop level.
The physical system in this case shows a non-Fermi liquid behavior. However
as doping is increased the \( FS \) immediately acquires curved sectors and
this opens up a possibility for Fermi liquid like behavior around certain regions
of \( \mathbf{k} \)-space. This feature seemed to be confirmed early on by
the ARPES data for the underdoped and optimally doped Bi2212 and YBCO compounds\cite{Shen}.
In the electronic spectra of these materials there appears an anisotropic pseudogap
and flat bands around \( \left( \pm \pi ,0\right)  \)and \( \left( 0,\pm \pi \right)  \)
and traces of gapless single-particle band dispersions around the \( \left( \pm \frac{\pi }{2},\pm \frac{\pi }{2}\right)  \)regions
of the Brillouin zone (BZ). This agrees qualitatively well with the phenomenological
picture of a FS composed of 'hot' and 'cold' spots put forwarded by Hlubina
and Rice and Pines and co-workers\cite{Hlubina}. In that picture the 'cold'
spots associated with correlated quasiparticle states are located along the
BZ diagonals. In contrast the 'hot' spots centered around \( \left( \pm \pi ,0\right)  \)and
\( \left( 0,\pm \pi \right)  \) are related to the pseudogap and other anomalies
of the cuprate normal phase. However recent photoemission experiments\cite{Valla}
which have a much better resolution than before put into doubt the applicability
of Fermi liquid theory even along the \( \left( 0,0\right)  \)- \( \left( \pi ,\pi \right)  \)
direction. Using their data on momentum widths as a function of temperature
for different points of FS, in optimally doped Bi2212, Valla et al show that
the imaginary part of the self-energy \( Im\Sigma  \) scales linearly with
the binding energy along that direction independent of the temperature. Similarly,
Kaminski et al show that the half-width-half-maximum of the spectral function
\( A\left( \mathbf{p},\omega \right)  \) single particle peak varies linearly
with \( \omega  \) above \( T_{c} \). They claim this to be analogous to both
the observed linear temperature behavior of the electrical resistivity and the
scattering rate. Those results are very different from those expected for a
Fermi liquid and support a marginal Fermi liquid phenomenology even near the
\( \left( \frac{\pi }{2},\frac{\pi }{2}\right)  \) points of the Brillouin
zone.

In this work we consider a a two-dimensional electron gas with a truncated \( FS \)
composed of four symmetric patches with both flat and conventionally curve arcs
in \( \mathbf{k} \)-space. These patches for simplicity are located around
\( \left( \pm k_{F}\right. \left. ,0\right)  \)and \( \left( 0,\pm k_{F}\right)  \)
respectively\( \left( Fig.1\right)  \). The Fermi liquid like states are defined
around the patch center. In contrast the border regions are taken to be flat.
As a result in this region the electron dispersion law is one-dimensional\cite{Ferraz1}.
In this way in each patch there are conventional two-dimensional electronic
states sandwiched by single-particles with a flat \( FS \) to simulate the
'cold' and 'hot' spots scenario described earlier on. Flat \( FS \) sectors
and single-particle with linear dispersion were also used earlier on by Dzyaloshinskii
and co-workers\cite{Dzyaloshinskii} to produce logarithmic singularities and
non-Fermi liquid behavior. Here they are used to test the stability of the two-dimensional
Fermi liquid states. We use the renormalization group \( \left( RG\right)  \)
method to deal with the infrared \( \left( IR\right)  \) singularities produced
in perturbation theory by the ``Cooper'', ``exchange'' and ``forward''channels.
Other \( RG \) methods were used recently by several workers to test the weak-coupling
limit of the two-dimensional Hubbard model with and without next-nearest neighbor
hopping against superconducting and magnetic ordering instabilities in different
doping regimes\cite{Zanchi}. However due to the difficulty in implementing
their method in higher orders they don't go beyond one-loop and no self-energy
effects are taken into account. As a result the coupling functions always have
divergent flows and there is never any sign of non-trivial fixed points.

The scope of this work is the following. We begin by reviewing briefly the model
used in our calculations. Next we calculate the one-particle irreducible functions
\( \Gamma ^{\left( 2\right) }_{\uparrow } \)and \( \Gamma ^{\left( 4\right) }_{\uparrow \downarrow ;\uparrow \downarrow } \)
up to two-loop order. We demonstrate that the quasiparticle weight \( Z \)
for the two-dimensional Fermi liquid state can vanish identically as a result
of the interaction of the ``cold'' particles with the flat sectors. We solve
the \( RG \) equation for the renormalized coupling in two-loop order and we
find a non-trivial IR stable fixed point. Later we estimate how higher-order
corrections and the mixing of the various scattering channels affect this result.
We calculate the spin and charge susceptibilities and discuss their physical
contents. We conclude by emphasizing that our results indicate the instability
of two -dimensional Fermi liquid states when they are renormalized by the interaction
with the flat sectors of the Fermi Surface and by arguing that they may well
be used to describe qualitatively the pseudogap phase of the cuprate superconductors.

\section{Two-Dimensional Model Fermi Surface}

Consider a \( 2d \) \( FS \) consisting of four disconnected patches centred
around \( \left( \pm k_{F}\right. \left. ,0\right)  \) and \( \left( 0,\right. \left. \pm k_{F}\right)  \).
Let us assume to begin with that they are Fermi liquid like. The disconnected
arcs separate occupied and unoccupied single-particle states along the direction
perpendicular to the Fermi Surface. However as we approach any patch along the
arc itself there is no sharp resolution of states in the vicinity of the gaps
located in the border regions. We assume that these regions are proper for non-Fermi
liquid \( \left( NFL\right)  \) behavior. To represent those \( NFL \) features
we take the \( FS \) to be flat in the border regions. In this way the single-particle
states which are a \( 2d \) Fermi liquid around the center of the patch acquire
an one-dimensional dispersion as we approach those flat border sectors. They
represent the ' hot' spots sandwiching the 'cold' spots in our model.

In order to be more quantitative consider the single-particle lagrangian density

\begin{equation}
\label{1}
\mathcal{L}=\sum _{\sigma }\psi ^{\dagger }_{\sigma }\left( x\right) \left( i\partial _{\overline{t}}+\frac{\nabla ^{2}}{2}+\overline{\varepsilon }_{F}\right) \psi _{\sigma }\left( x\right) -U\psi ^{\dagger }_{\uparrow }\left( x\right) \psi ^{\dagger }_{\downarrow }\left( x\right) \psi _{\downarrow }\left( x\right) \psi _{\uparrow }\left( x\right) 
\end{equation}
where \( x=\left( \overline{t},\mathbf{x}\right)  \),\( \overline{\varepsilon }_{F}=k^{2}_{F}/2 \),
\( \overline{t}^{-1}=m^{*}t^{-1} \) with \( m^{*} \) being the effective mass.
When we proceed with our renormalization scheme in the vicinity of a given \( FS \)
point we replace \( k_{F} \) by the corresponding bare \( k_{F}^{0}=Z^{-1}\Lambda \overline{k}_{F} \)
where \( \overline{k}_{F} \) is dimensionless with \( \Lambda  \) being a
momentum scale. In this way \( \overline{k}_{F} \) can be finite even if both
\( Z \) and \( \Lambda \rightarrow 0 \) and the coupling constant \( U \)
scales with momenta as \( \Lambda ^{2-d} \) in \( d \) spatial dimensions.
Here the fermion fields are non-zero only in a slab of width \( 2\lambda  \)
around the four symmetric patches of \( FS \). Thus in momentum space the single-particle
\( \varepsilon \left( \mathbf{p}\right)  \) is defined according to the sector
and patch under consideration. For example,in the vicinity of the central zone
of the patch defined around the \( FS \) point \( \left( 0,-k_{F}\right)  \)
we have that

\begin{equation}
\label{2}
\varepsilon \left( \mathbf{p}\right) \cong \frac{k_{F}^{2}}{2}-k_{F}\left( p_{y}+k_{F}\right) +\frac{p_{x}^{2}}{2},
\end{equation}
with \( -\Delta \leq p_{x}\leq \Delta  \) . In contrast in the border regions
of the same patch we find instead

\begin{equation}
\label{3}
\varepsilon \left( \mathbf{p}\right) \cong \frac{k_{F}^{2}}{2}-k_{F}\left( p_{y}+k_{F}-\frac{\Delta ^{2}}{2k_{F}}\right) ,
\end{equation}
for \( \Delta \leq p_{x}\leq \lambda  \) or \( -\lambda \leq p_{x}\leq -\Delta  \)
. We follow the same scheme to define \( \varepsilon \left( \mathbf{p}\right)  \)
in all other patches of \( FS \).

In setting up our perturbation theory scheme two quantities appear frequently:
the particle-hole and the particle-particle bubble diagrams. In zero-th order
they are defined respectively as

\begin{equation}
\label{4}
\chi ^{\left( 0\right) }_{\uparrow \downarrow }\left( P\right) =-\int _{q}G^{\left( 0\right) }_{\uparrow }\left( q\right) G_{\downarrow }^{\left( 0\right) }\left( q+P\right) ,
\end{equation}
and

\begin{equation}
\label{5}
\Pi ^{\left( 0\right) }_{\uparrow \downarrow }\left( P\right) =\int _{q}G^{\left( 0\right) }_{\uparrow }\left( q\right) G^{\left( 0\right) }_{\downarrow }\left( -q+P\right) 
\end{equation}
where 
\begin{equation}
\label{6}
G^{\left( 0\right) }_{\uparrow }\left( q_{0},\mathbf{q}\right) =\frac{\theta \left( \overline{\varepsilon }\left( \mathbf{q}\right) \right) }{q_{0}-\overline{\varepsilon }\left( \mathbf{q}\right) +i\delta }+\frac{\theta \left( -\overline{\varepsilon }\left( \mathbf{q}\right) \right) }{q_{0}-\overline{\varepsilon }\left( \mathbf{q}\right) -i\delta }
\end{equation}
with \( \overline{\varepsilon }\left( \mathbf{q}\right) =\varepsilon \left( \mathbf{q}\right) -\frac{k^{2}_{F}}{2} \),
\( \int _{q}=-i\int \frac{dq_{0}}{2\pi }\int \frac{d\mathbf{q}}{\left( 2\pi \right) ^{2}} \)and
\( q=\left( q_{0},\mathbf{q}\right)  \).

It turns out that \( \chi ^{\left( 0\right) } \) is singular only if the \( G^{\left( o\right) } \)'s
refer to flat sectors such that \( \mathbf{q} \) and \( \mathbf{q}+\mathbf{P} \)
are points from corresponding antipodal border regions of \( FS \). In the
case, in which e.g. \( \mathbf{P}=\left( 0,2k_{F}-\frac{\Delta ^{2}}{k_{F}}\right)  \)
we find

\begin{equation}
\label{7}
\chi ^{\left( 0\right) }_{\uparrow \downarrow }\left( \mathbf{P};P_{0}\right) =\frac{\left( \lambda -\Delta \right) }{4\pi ^{2}k_{F}}\left[ \ln \left( \frac{\Omega +P_{0}-i\delta }{P_{0}-i\delta }\right) +\ln \left( \frac{\Omega -P_{0}-i\delta }{-P_{0}-i\delta }\right) \right] 
\end{equation}
with \( \Omega =2k_{F}\lambda  \).

In contrast \( \Pi ^{\left( 0\right) } \)is singular for particles located
in both 'cold' and 'hot' spots whenever they are involved in a Cooper scattering
channel. Here for e.g. \( \mathbf{P}=\left( 0,0\right)  \) we obtain

\begin{equation}
\label{8}
\Pi ^{\left( 0\right) }_{\uparrow \downarrow }\left( \mathbf{P};P_{0}\right) =\frac{\lambda }{\pi ^{2}k_{F}}\left[ \ln \left( \frac{\Omega +P_{0}-i\delta }{P_{0}-i\delta }\right) +\ln \left( \frac{\Omega -P_{0}-i\delta }{-P_{0}-i\delta }\right) \right] 
\end{equation}
As is well known the Cooper channel singularity drives the system towards its
superconducting instability. However at one-loop for a repulsive interactions
the renormalized coupling approaches the trivial Fermi liquid fixed point\cite{Polchinski}.
As opposed to that the singularity in \( \chi ^{\left( 0\right) } \) produced
by the one-loop exchange channel drives the physical system to a non-perturbative
regime. This non-perturbative behavior might be indicative of either the failure
of the one-loop truncation or of the inadequacy of perturbation theory itself
to deal with that situation. To find out what is in fact the case we consider
the effect of higher-order contributions in both one-particle irreducible functions
\( \Gamma ^{\left( 2\right) }\left( p\right)  \) and \( \Gamma ^{\left( 4\right) }\left( p\right)  \).

\section{One-Particle Irreducible Functions }

Let us initially consider the one-particle irreducible function \( \Gamma ^{\left( 2\right) }_{\uparrow }\left( p_{0},\mathbf{p}\right)  \)
for a \( \mathbf{p} \) located in the vicinity of a 'cold' spot point of \( FS \)
such as \( \mathbf{p}^{*}=\left( \Delta ,-k_{F}+\frac{\Delta ^{2}}{2k_{F}}\right)  \).
We can write \( \Gamma ^{\left( 2\right) } \)in this case as

\begin{equation}
\label{9}
\Gamma ^{\left( 2\right) }_{\uparrow }\left( p_{0},\mathbf{p}\right) =p_{0}+k_{F}\left( p_{y}+k_{F}-\frac{\Delta ^{2}}{k_{F}}\right) -\Sigma _{\uparrow }\left( p_{0},\mathbf{p}\right) 
\end{equation}
where, using perturbation theory, the two-loop self-energy \( \Sigma _{\uparrow } \)
is given by

\begin{equation}
\label{10}
\Sigma _{\uparrow }\left( p_{0},\mathbf{p}\right) =\frac{2U\lambda ^{2}}{\pi ^{2}}-2U^{2}\int _{q}G^{\left( 0\right) }_{\downarrow }\left( q\right) \chi ^{\left( 0\right) }_{\uparrow \downarrow }\left( q-p\right) 
\end{equation}
The constant term produces at this order a constant shift in \( k_{F} \) which
will be neglected from now on. Evaluating the integrals over \( q \) we obtain\cite{Ferraz2}

\begin{eqnarray}
\Sigma _{\uparrow }\left( p_{0},\mathbf{p}\right) \cong -\frac{3U^{2}}{64\pi ^{4}}\left( \frac{\lambda -\Delta }{k_{F}}\right) ^{2}\left( p_{0}+k_{F}\left( p_{y}+k_{F}-\frac{\Delta ^{2}}{2k_{F}}\right) \right)  &  & \\
.\left( \ln \left( \frac{\Omega +p_{0}-i\delta }{k_{F}\left( p_{y}+k_{F}-\frac{\Delta ^{2}}{2k_{F}}\right) +p_{0}-i\delta }\right) \right.  &  & \\
+\left. \ln \left( \frac{\Omega -p_{0}-i\delta }{k_{F}\left( p_{y}+k_{F}-\frac{\Delta ^{2}}{2k_{F}}\right) -p_{0}-i\delta }\right) \right)  &  & \label{11} 
\end{eqnarray}

Clearly both \( \partial \Sigma _{\uparrow }/\partial p_{y} \) and \( \partial \Sigma _{\uparrow }/\partial p_{0} \)
are divergent at \( FS \). This gives the marginal Fermi liquid result \cite{Varma}
for \( p_{y}=-k_{F}+\frac{\Delta ^{2}}{2k_{F}} \) and \( p_{0}\rightarrow 0 \)
which nullifies the quasiparticle weight \( Z=1-\partial Re\Sigma _{\uparrow }/\partial p_{0}\left. \right| _{\mathbf{p}^{*};\omega } \)
at the Fermi Surface. We can also arrive at this result using the renormalization
group \( \left( RG\right)  \). For this we define the renormalized one-particle
irreducible function \( \Gamma ^{\left( 2\right) }_{R\uparrow }\left( p_{0},\mathbf{p}\right)  \)
such that at \( p_{0}=\omega  \), where \( \omega  \) is a small energy scale
parameter, and \( \mathbf{p}=\mathbf{p}^{*} \), \( \Gamma ^{\left( 2\right) }_{R\uparrow }\left( p_{0}=\omega ,\mathbf{p}=\mathbf{p}^{*}\right) =\omega  \).
Using \( RG \), \( \Gamma ^{\left( 2\right) }_{R\uparrow } \) is related to
the corresponding bare function \( \Gamma ^{\left( 2\right) }_{0\uparrow } \)
by

\begin{equation}
\label{12}
\Gamma ^{\left( 2\right) }_{R\uparrow }\left( p;U;\omega \right) =Z\left( \mathbf{p}^{*};\omega \right) \Gamma ^{\left( 2\right) }_{0\uparrow }\left( p;U_{0}\right) 
\end{equation}
where \( U_{0} \) is the corresponding bare coupling. Since at zero-th order
\( U_{0}=U \) it follows from our prescription and perturbative result that

\begin{equation}
\label{13}
Z\left( \mathbf{p}^{*};\omega \right) =\frac{1}{1+\frac{3U^{2}}{32\pi ^{4}}\left( \frac{\lambda -\Delta }{k_{F}}\right) ^{2}\ln \left( \frac{\Omega }{\omega }\right) }
\end{equation}
Naturally, \( Z=0 \) if \( \omega \rightarrow 0 \). As we showed elsewhere
this result reflects itself in the anomalous dimension developed by the single-particle
Green's function at \( FS \). 

Let us next calculate the one-particle irreducible two-particle function \( \Gamma _{\alpha ,\beta }^{\left( 4\right) }\left( p_{1},p_{2};p_{3},p_{4}\right)  \)
for \( \alpha ,\beta =\uparrow ,\downarrow  \). This function depends on the
spin arrangements of the external legs as well as on the scattering channel
into consideration. Generically for antiparallel spins up to two-loop order
we have that

\begin{eqnarray}
\Gamma ^{\left( 4\right) }_{\uparrow \downarrow }\left( p_{1},p_{2};p_{3},p_{4}\right) =-U+U^{2}\int _{k}G^{\left( 0\right) }_{\uparrow }\left( k\right) G^{\left( 0\right) }_{\downarrow }\left( k+p_{4}-p_{1}\right)  &  & \\
+U^{2}\int _{k}G^{\left( 0\right) }_{\uparrow }\left( k\right) G^{\left( 0\right) }_{\downarrow }\left( -k+p_{1}+p_{2}\right) + &  & \\
-U^{3}\int _{k}G^{\left( 0\right) }_{\downarrow }\left( k\right) G^{\left( 0\right) }_{\downarrow }\left( k+p_{3}-p_{1}\right) \int _{q}G^{\left( 0\right) }_{\uparrow }\left( q\right) G^{\left( 0\right) }_{\uparrow }\left( q+p_{3}-p_{1}\right) +... & \label{14} 
\end{eqnarray}

The only one-loop terms are the particle-hole diagram which couples legs with
opposite spins and the particle-particle diagram for legs with opposite spins.
The forward-channel which is associated with diagrams with external legs of
the same side with the same spin only begins contributes to \( \Gamma ^{\left( 4\right) }_{\uparrow \downarrow } \)
from the two-loop order on. The other two-loop contributions are omitted here
for economy of space. In contrast, if we now consider the two-particle function
for parallel spins we find instead

\begin{eqnarray}
\Gamma ^{\left( 4\right) }_{\uparrow \uparrow }\left( p_{1},p_{2};p_{3},p_{4}\right) =-U^{2}\int _{k}G_{\downarrow }^{\left( 0\right) }\left( k\right) G^{\left( 0\right) }_{\downarrow }\left( k+p_{3}-p_{1}\right) + &  & \\
+U^{3}\int _{k}G_{\downarrow }^{\left( 0\right) }\left( k\right) G^{\left( 0\right) }_{\downarrow }\left( k+p_{3}-p_{1}\right) \Pi ^{\left( 0\right) }_{\uparrow \downarrow }\left( k+p_{2}\right) + &  & \label{15} \\
-U^{3}\int _{k}G_{\downarrow }^{\left( 0\right) }\left( k\right) G^{\left( 0\right) }_{\downarrow }\left( k+p_{3}-p_{1}\right) \chi ^{\left( 0\right) }_{\uparrow \downarrow }\left( p_{4}-k\right) +... &  & 
\end{eqnarray}
 Clearly the singularities in our perturbation series expansions depend very
crucially on the values of the external momenta. Due to this anisotropy of momenta
space different scattering channels produce different divergent results at different
positions of \( FS \). As we will see later this automatically requires the
definition of momenta dependent bare coupling functions in our perturbation
series expansions. Despite that all the existing divergences can be grouped
together with respect with their scattering channel and vertex type associated
with the singular loop integrations. This opens the way to define a systematic
regularization procedure to guide our renormalization group prescriptions throughout
all our calculations. Here we follow the convention to define the ``exchange''
type vertex when its left \( \left( right\right)  \) incoming and outgoing
lines have opposite spins. We have an ``exchange'' type divergence whenever
the momenta of particles with opposite spins can be tunned together to produce
a logarithmic singularity in the one-loop particle-hole diagram. This can be
easily achieved in the `` exchange'' scattering channel for \( \uparrow \mathbf{p}_{1}=\uparrow \mathbf{p}_{3} \)
and \( \downarrow \mathbf{p}_{2}=\downarrow \mathbf{p}_{4} \). In contrast
for \( \uparrow \mathbf{p}_{1}=-\mathbf{p}_{2}\downarrow  \) and \( \uparrow \mathbf{p}_{3}=-\downarrow \mathbf{p}_{4} \)
it is now the particle-particle diagram which gives the leading contribution
in one-loop order. When this is the case we say we have a ``Cooper `` channel.
Now the divergent particle-particle loop can be described in terms of bare ``Cooper'
vertices. Finally we say we have bare ``forward'' vertex whenever a particle-hole
loop produced by left \( \left( right\right)  \) incoming and outgoing particles
with same spins becomes divergent. This vertices appear naturally in the ``forward''
( zero sound )channel for \( \uparrow \mathbf{p}_{1}=\downarrow \mathbf{p}_{4} \)
and \( \downarrow \mathbf{p}_{2}=\uparrow \mathbf{p}_{3} \). Here it is now
\( \uparrow \mathbf{p}_{3} \) and \( \uparrow \mathbf{p}_{1} \) which are
tunned to produce a \( \ln ^{2} \) divergence in two-loop order. In this work
we consider only the leading divergence at every order of perturbation theory.
Nevertheless since we go beyond one-loop and we include non-trivial self-energy
corrections we take explicitly into account contributions which don't appear
either in parquet type or numerical \( RG \) approaches. Inasmuch as both the
renormalization conditions and the bare coupling functions vary as we move along
in momenta space, strictly speaking, we need an infinite number of counter-terms
to regularize our model. However all the divergences which appear in perturbation
theory can be associated with a loop integration with vertices which are either
of exchange, Cooper or forward type. As we will show explicitly later it is
possible to define three bare coupling functions \( U_{0x}\left( \downarrow \mathbf{p}_{4}-\uparrow \mathbf{p}_{1}\right)  \),
\( U_{0C}\left( \uparrow \mathbf{p}_{1}+\mathbf{p}_{2}\downarrow \right)  \)
and \( U_{0f}\left( \uparrow \mathbf{p}_{3}-\uparrow \mathbf{p}_{1}\right)  \)
respectively to cancel out exactly, order by order, all the divergences which
appear in our perturbation theory expansions.

To illustrate our argument further take initially e.g. \( \mathbf{p}_{1}=\mathbf{p}_{3}=\left( \Delta ,k_{F}-\frac{\Delta ^{2}}{2k_{F}}\right)  \)
and \( \mathbf{p}_{2}=\mathbf{p}_{4}=\left( \lambda -\epsilon ,-k_{F}+\frac{\Delta ^{2}}{2k_{F}}\right)  \)
with \( \epsilon  \) being such that \( 0\leq \epsilon <\lambda -\Delta  \).
The leading terms up to two-loop order for \( p_{0}\approx 0 \) are \( \left( Fig.2a\right)  \)

\begin{eqnarray}
\Gamma ^{\left( 4\right) }_{\uparrow \downarrow }\left( \mathbf{p}_{1}=\mathbf{p}_{3};\mathbf{p}_{2}=\mathbf{p}_{4},p_{0}\right) =-U-U^{2}\chi ^{\left( 0\right) }_{\uparrow \downarrow }\left( \mathbf{p}_{4}-\mathbf{p}_{1};p_{0}\right) +U^{2}\Pi ^{\left( 0\right) }_{\uparrow \downarrow }\left( \mathbf{p}_{1}+\mathbf{p}_{2};p_{0}\right)  &  & \\
-U^{3}\left( \chi _{\uparrow \downarrow }^{\left( 0\right) }\left( \mathbf{p}_{4}-\mathbf{p}_{1};p_{0}\right) \right) ^{2}-U^{3}\left( \Pi ^{\left( 0\right) }\left( \mathbf{p}_{1}+\mathbf{p}_{2};p_{0}\right) \right) ^{2} &  & \\
-U^{3}\int _{k}G^{\left( 0\right) }_{\uparrow }\left( k\right) G_{\downarrow }^{\left( 0\right) }\left( k+p_{4}-p_{1}\right) \Pi ^{\left( 0\right) }_{\uparrow \downarrow }\left( k+p_{2}\right)  &  & \\
+U^{3}\int _{k}G^{\left( 0\right) }_{\uparrow }\left( k\right) G_{\downarrow }^{\left( 0\right) }\left( -k+p_{1}+p_{2}\right) \left( \chi ^{\left( 0\right) }_{\uparrow \downarrow }\left( p_{4}-k\right) +p_{4}\rightleftharpoons p_{3}\right) +... &  & \label{22} 
\end{eqnarray}

If we evaluate all diagrams we find

\begin{eqnarray}
\Gamma ^{\left( 4\right) }_{\uparrow \downarrow }\left( \mathbf{p}_{1}=\mathbf{p}_{3};\mathbf{p}_{2}=\mathbf{p}_{4},p_{0}\approx \omega \right) =-U-\frac{U^{2}}{2\pi ^{2}k_{F}}\epsilon \ln \left( \frac{\Omega }{\omega }\right)  &  & \\
+\frac{U^{2}}{2\pi ^{2}k_{F}}\left( \lambda -\Delta -\epsilon \right) \ln \left( \frac{\Omega }{\omega }\right) -\frac{U^{3}}{4\pi ^{4}k_{F}^{2}}\epsilon ^{2}\left( \ln \left( \frac{\Omega }{\omega }\right) \right) ^{2} &  & \\
-\frac{U^{3}}{4\pi ^{4}k_{F}^{2}}\left( \lambda -\Delta -\epsilon \right) ^{2}\left( \ln \left( \frac{\Omega }{\omega }\right) \right) ^{2}+\frac{U^{3}}{16\pi ^{4}k_{F}^{2}}\left[ 3\epsilon \left( \lambda -\Delta -\frac{\epsilon }{2}\right) \right.  &  & \label{14} \\
\left. +\left( \left( \lambda -\Delta \right) ^{2}-\epsilon ^{2}\right) \right] \left( \ln \left( \frac{\Omega }{\omega }\right) \right) ^{2}+... &  & 
\end{eqnarray}

For \( \mathbf{p}_{1}=-\mathbf{p}_{2}=\left( \Delta ,k_{F}-\frac{\Delta ^{2}}{2k_{F}}\right)  \)
and \( \mathbf{p}_{3}=-\mathbf{p}_{4}=\left( \lambda -\epsilon ,-k_{F}+\frac{\Delta ^{2}}{2k_{F}}\right)  \)
up to two-loop order our series expansion becomes instead \( \left( Fig.2b\right)  \)

\begin{eqnarray}
\Gamma ^{\left( 4\right) }_{\uparrow \downarrow }\left( \mathbf{p}_{1}=-\mathbf{p}_{2};p_{0}\right) =-U+U^{2}\Pi ^{\left( 0\right) }_{\uparrow \downarrow }\left( p_{0}\right) -U^{2}\chi ^{\left( 0\right) }_{\uparrow \downarrow }\left( \mathbf{p}_{4}-\mathbf{p}_{1};p_{0}\right)  &  & \\
-U^{3}\left( \Pi ^{\left( 0\right) }_{\uparrow \downarrow }\left( p_{0}\right) \right) ^{2}-U^{3}\left( \chi _{\uparrow \downarrow }^{\left( 0\right) }\left( \mathbf{p}_{4}-\mathbf{p}_{1};p_{0}\right) \right) ^{2} &  & \\
+U^{3}\int _{k}G^{\left( 0\right) }_{\downarrow }\left( k\right) G^{\left( 0\right) }_{\uparrow }\left( -k;p_{0}\right) \left( \chi ^{\left( 0\right) }_{\uparrow \downarrow }\left( p_{3}-k\right) +\chi ^{\left( 0\right) }_{\uparrow \downarrow }\left( p_{4}-k\right) \right)  &  & \label{15} \\
-U^{3}\int _{k}G^{\left( 0\right) }_{\uparrow }\left( k\right) G_{\downarrow }^{\left( 0\right) }\left( k+p_{4}-p_{1}\right) \Pi ^{\left( 0\right) }\left( k+p_{2}\right) +... &  & 
\end{eqnarray}
Evaluating the integrals we obtain

\begin{eqnarray}
\Gamma ^{\left( 4\right) }_{\uparrow \downarrow }\left( \mathbf{p}_{1}=-\mathbf{p}_{2};p_{0}\right) =-U+\frac{U^{2}}{2\pi ^{2}k_{F}}\left( 4\lambda \right) \ln \left( \frac{\Omega }{\omega }\right)  &  & \\
-\frac{U^{2}}{4\pi ^{2}k_{F}}\left( \lambda -\Delta -\epsilon \right) \ln \left( \frac{\Omega }{\omega }\right) -\frac{U^{3}}{4\pi ^{4}k_{F}^{2}}\left( 4\lambda \right) ^{2}\left( \ln \left( \frac{\Omega }{\omega }\right) \right) ^{2} &  & \\
-\frac{U^{3}}{16\pi ^{4}k_{F}^{2}}\left( \lambda -\Delta -\epsilon \right) ^{2}\left( \ln \left( \frac{\Omega }{\omega }\right) \right) ^{2} &  & \\
+\frac{U^{3}}{16\pi ^{4}k_{F}^{2}}\left[ \frac{3}{2}\left( \lambda -\Delta \right) ^{2}+2\epsilon \left( \lambda -\Delta -\epsilon \right) \right] \left( \ln \left( \frac{\Omega }{\omega }\right) \right) ^{2} &  & \\
+\frac{U^{3}}{16\pi ^{4}k_{F}^{2}}\left[ \left( \lambda -\Delta \right) ^{2}-\epsilon ^{2}\right] \left( \ln \left( \frac{\Omega }{\omega }\right) \right) ^{2} & +... & \label{16} 
\end{eqnarray}

Finally, for \( \mathbf{p}_{1}=\mathbf{p}_{4}=\left( \Delta ,k_{F}-\frac{\Delta ^{2}}{2k_{F}}\right)  \)
and \( \mathbf{p}_{2}=\mathbf{p}_{3}=\left( \lambda -\epsilon ,-k_{F}+\frac{\Delta ^{2}}{2k_{F}}\right)  \)we
have that \( \mathbf{p}_{3}-\mathbf{p}_{1}=\left( \lambda -\Delta -\epsilon ,-2k_{F}+\frac{\Delta ^{2}}{k_{F}}\right)  \),
\( \mathbf{p}_{1}+\mathbf{p}_{2}=\left( \lambda +\Delta -\epsilon \right)  \)
and our series expansion in the forward channel becomes \( \left( Fig.2c\right)  \)

\begin{eqnarray}
\Gamma ^{\left( 4\right) }_{\uparrow \downarrow }\left( \mathbf{p}_{1}=\mathbf{p}_{4};\mathbf{p}_{2}=\mathbf{p}_{3},p_{0}\right) =-U+U^{2}\Pi ^{\left( 0\right) }\left( \mathbf{p}_{1}+\mathbf{p}_{2};p_{0}\right)  &  & \\
-U^{3}\left( \chi _{\uparrow \downarrow }^{\left( 0\right) }\left( \mathbf{p}_{3}-\mathbf{p}_{1};p_{0}\right) \right) ^{2}-U^{3}\left( \Pi ^{\left( 0\right) }\left( \mathbf{p}_{1}+\mathbf{p}_{2};p_{0}\right) \right) ^{2} &  & \\
+U^{3}\int _{k}G^{\left( 0\right) }_{\uparrow }\left( k\right) G_{\downarrow }^{\left( 0\right) }\left( -k+p_{1}+p_{2}\right) \left( \chi ^{\left( 0\right) }_{\uparrow \downarrow }\left( p_{4}-k\right) +p_{4}\rightleftharpoons p_{3}\right)  &  & \\
+U^{3}\int _{k}G^{\left( 0\right) }_{\downarrow }\left( k+p_{3}-p_{1}\right) G^{\left( 0\right) }_{\uparrow }\left( k\right) \chi ^{\left( 0\right) }_{\uparrow \downarrow }\left( p_{4}-k\right) +... &  & \label{16} 
\end{eqnarray}
Solving all the integrals above we get

\begin{eqnarray}
\Gamma ^{\left( 4\right) }_{\uparrow \downarrow }\left( \mathbf{p}_{1}=\mathbf{p}_{4};\mathbf{p}_{2}=\mathbf{p}_{3},p_{0}\right) =-U & +\frac{U^{2}}{2\pi ^{2}k_{F}}\left( \lambda -\Delta -\epsilon \right) \ln \left( \frac{\Omega }{\omega }\right)  & \label{17} \\
-\frac{U^{3}}{16\pi ^{4}k_{F}^{2}}\left( \epsilon \right) ^{2}\left( \ln \left( \frac{\Omega }{\omega }\right) \right)  & -\frac{U^{3}}{16\pi ^{4}k_{F}^{2}}\left( \lambda -\Delta -\epsilon \right) ^{2}\left( \ln \left( \frac{\Omega }{\omega }\right) \right) ^{2} & \\
 & +\frac{U^{3}}{16\pi ^{4}k_{F}^{2}}\left[ \left( \lambda -\Delta \right) ^{2}-\epsilon ^{2}\right] \left( \ln \left( \frac{\Omega }{\omega }\right) \right) ^{2}+...
\end{eqnarray}

Our renormalization prescription must therefore incorporate this momenta space
anisotropy to cancel all the corresponding singularities appropriately. Using
\( RG \) theory we now define the renormalized two-particle function \( \Gamma ^{\left( 4\right) }_{R\uparrow \downarrow } \)in
terms of the corresponding bare function \( \Gamma _{0\uparrow \downarrow }^{\left( 4\right) } \)
:

\begin{equation}
\label{15}
\Gamma ^{\left( 4\right) }_{R\uparrow \downarrow }\left( p_{1},p_{2};p_{3},p_{4};U_{a}\left( \left\{ \mathbf{p}_{i}\right\} ;\omega \right) ;\omega \right) =\prod _{i=1}^{4}Z^{\frac{1}{2}}\left( \mathbf{p}_{i};\omega \right) \Gamma ^{\left( 4\right) }_{0\uparrow \downarrow }\left( p_{1},p_{2};p_{3},p_{4};U_{0a}\left( \mathbf{p}_{\mathbf{i}}\right) \right) 
\end{equation}
 where \( U_{0a} \) is the bare coupling and \( U_{a} \) is the corresponding
renormalized coupling with \( a=x\left( exchange\right) ;C\left( Cooper\right) ;f\left( forward\right)  \).
We say we are in an exchange type channel whenever the exchange particle-hole
diagram gives the leading contribution in first order perturbation theory. Similarly
for the Cooper channel the dominant first order contribution is the Cooper particle-particle
loop. Finally when the particle-hole loop with left and right legs with the
same spin is divergent we are in the forward channel. Using this scheme for
the momenta values above in the exchange channel the corresponding renormalized
coupling is determined by the prescription

\begin{equation}
\label{16}
\Gamma ^{\left( 4\right) }_{R\uparrow \downarrow }\left( \mathbf{p}_{1}=\mathbf{p}_{3};\mathbf{p}_{2}=\mathbf{p}_{4};p_{0}=\omega ;U_{a}\right) =-U_{x}\left( \mathbf{p}_{1}=\mathbf{p}_{3};\mathbf{p}_{2}=\mathbf{p}_{4};\omega ;U_{a}\right) 
\end{equation}
for \( \mathbf{p}_{1}=\mathbf{p}_{3}=\left( \Delta ,k_{F}-\frac{\Delta ^{2}}{2k_{F}}\right)  \)
and \( \mathbf{p}_{2}=\mathbf{p}_{4}=\left( \lambda -\epsilon ,-k_{F}+\frac{\Delta ^{2}}{2k_{F}}\right)  \)
with \( 2\epsilon >\lambda -\Delta  \).

Similarly, in the Cooper channel we assume that the renormalized coupling is
fixed by

\begin{equation}
\label{17}
\Gamma ^{\left( 4\right) }_{R\uparrow \downarrow }\left( \mathbf{p}_{1}=-\mathbf{p}_{2};p_{0}=\omega ;U_{a}\right) =-U_{C}\left( \mathbf{p}_{1}=-\mathbf{p}_{2};\omega ;U_{a}\right) 
\end{equation}
for \( \mathbf{p}_{1}=-\mathbf{p}_{2}=\left( \Delta ,k_{F}-\frac{\Delta ^{2}}{2k_{F}}\right)  \)
and \( \mathbf{p}_{1}=-\mathbf{p}_{2}=\left( \Delta ,k_{F}-\frac{\Delta ^{2}}{2k_{F}}\right)  \).
Finally for the forward channel we define

\begin{equation}
\label{18}
\Gamma ^{\left( 4\right) }_{R\uparrow \downarrow }\left( \mathbf{p}_{1}=\mathbf{p}_{4};\mathbf{p}_{2}=\mathbf{p}_{3},p_{0}=\omega ;U_{a}\right) =-U_{f}\left( \mathbf{p}_{1}=\mathbf{p}_{4};\mathbf{p}_{2}=\mathbf{p}_{3};\omega ;U_{a}\right) 
\end{equation}
for \( \mathbf{p}_{1}=\mathbf{p}_{4}=\left( \Delta ,k_{F}-\frac{\Delta ^{2}}{2k_{F}}\right)  \)
and \( \mathbf{p}_{2}=\mathbf{p}_{3}=\left( \lambda -\epsilon ,-k_{F}+\frac{\Delta ^{2}}{2k_{F}}\right)  \)
respectively. Using these prescriptions we find respectively

\begin{eqnarray}
U_{x}\left( \mathbf{p}_{1}=\mathbf{p}_{3};\mathbf{p}_{2}=\mathbf{p}_{4};\omega ;U_{a}\right) =Z\left( \mathbf{p}_{1};\omega \right) Z\left( \mathbf{p}_{4};\omega \right) \left\{ U_{0x}\left[ 1+\right. \right.  &  & \\
+\frac{\epsilon }{2\pi ^{2}k_{F}}U_{0x}\ln \left( \frac{\Omega }{\omega }\right) +\frac{1}{4\pi ^{2}k^{2}_{F}}\left( \epsilon ^{2}U_{0x}^{2}+\right.  &  & \\
\left. \left. -\frac{3}{4}\epsilon \left( \lambda -\Delta -\frac{\epsilon }{2}\right) U^{2}_{0C}\right) \left( \ln \left( \frac{\Omega }{\omega }\right) \right) ^{2}+...\right] + &  & \\
-\frac{\lambda -\Delta -\epsilon }{2\pi ^{2}k_{F}}U^{2}_{0C}\ln \left( \frac{\Omega }{\omega }\right) +\frac{U_{0C}}{4\pi ^{4}k^{2}_{F}}\left[ \left( \lambda -\Delta -\epsilon \right) ^{2}U^{2}_{0C}+\right.  &  & \\
\left. \left. -\frac{1}{4}\left( \left( \lambda -\Delta \right) ^{2}-\epsilon ^{2}\right) U^{2}_{0x}\right] \left( \ln \left( \frac{\Omega }{\omega }\right) \right) ^{2}+...\right\}  &  & \label{19} 
\end{eqnarray}
for the exchange channel,

\begin{eqnarray}
U_{C}\left( \mathbf{p}_{1}=-\mathbf{p}_{2};\mathbf{p}_{3}=-\mathbf{p}_{4};\omega ;U_{a}\right) =Z\left( \mathbf{p}_{1};\omega \right) Z\left( \mathbf{p}_{4};\omega \right) \left\{ U_{0C}\left[ 1+\right. \right.  &  & \\
-\frac{4\lambda }{2\pi ^{2}k_{F}}U_{0C}\ln \left( \frac{\Omega }{\omega }\right) +\frac{1}{4\pi ^{2}k^{2}_{F}}\left( \left( 4\lambda \right) ^{2}U_{0C}^{2}+\right.  &  & \\
\left. \left. -\frac{1}{4}\left( \frac{3}{2}\left( \lambda -\Delta \right) ^{2}+2\epsilon \left( \lambda -\Delta -\epsilon \right) \right) U^{2}_{0x}\right) \left( \ln \left( \frac{\Omega }{\omega }\right) \right) ^{2}+...\right] + &  & \\
+\frac{\lambda -\Delta -\epsilon }{4\pi ^{2}k_{F}}U^{2}_{0x}\ln \left( \frac{\Omega }{\omega }\right) +\frac{U_{0x}}{16\pi ^{4}k^{2}_{F}}\left[ \left( \lambda -\Delta -\epsilon \right) ^{2}U^{2}_{0x}+\right.  &  & \\
\left. \left. -\left( \left( \lambda -\Delta \right) ^{2}-\epsilon ^{2}\right) U^{2}_{0C}\right] \left( \ln \left( \frac{\Omega }{\omega }\right) \right) ^{2}+...\right\}  &  & \label{20} 
\end{eqnarray}
for the Cooper channel and finally

\begin{eqnarray}
U_{f}\left( \mathbf{p}_{1}=\mathbf{p}_{4};\mathbf{p}_{2}=\mathbf{p}_{3};\omega ;U_{a}\right) =Z\left( \mathbf{p}_{1};\omega \right) Z\left( \mathbf{p}_{3};\omega \right) \left\{ U_{0f}\left[ 1+\right. \right.  &  & \\
\left. +\frac{\epsilon ^{2}}{4\pi ^{4}k^{2}_{F}}U^{2}_{0f}\ln ^{2}\left( \frac{\Omega }{\omega }\right) +...\right] -\frac{\lambda -\Delta -\epsilon }{2\pi ^{2}k_{F}}U^{2}_{0C}\ln \left( \frac{\Omega }{\omega }\right) + &  & \\
\left. +\frac{U_{0C}}{4\pi ^{4}k^{2}_{F}}\left( \left( \lambda -\Delta -\epsilon \right) ^{2}U^{2}_{0C}-\frac{1}{4}\left[ \left( \lambda -\Delta \right) ^{2}-\epsilon ^{2}\right] U^{2}_{0x}\right) \left( \ln \left( \frac{\Omega }{\omega }\right) \right) ^{2}+...\right\}  &  & \label{21} 
\end{eqnarray}
for the forward channel with \( Z\left( \mathbf{p}_{1};\omega \right)  \) given
as before and

\begin{equation}
\label{22}
Z^{-1}\left( \mathbf{p}_{3};\omega \right) =1+\frac{U^{2}_{0x}}{16\pi ^{4}k^{2}_{F}}\left[ \frac{3}{2}\left( \lambda -\Delta \right) ^{2}+3\epsilon \left( \lambda -\Delta -\epsilon \right) \right] \ln \left( \frac{\Omega }{\omega }\right) +...
\end{equation}

However taking into consideration that the divergencies are removed by local
subtractions and at third-order perturbation theory we must have \( U^{2}_{0C}\cong U^{2}_{0x}\cong U^{2}_{0f}\cong U_{0}^{2} \)
there is no mixing of channels if we only do perturbation theory up to two-loop
order. As a result the renormalized couplings reduce to

\begin{equation}
\label{17}
U_{x}=U_{0x}-\left( aU_{0x}^{2}-2bU_{0x}^{3}\right) \ln \left( \frac{\Omega }{\omega }\right) +...,
\end{equation}
where

\begin{eqnarray}
a=\frac{2\epsilon -\left( \lambda -\Delta \right) }{2\pi ^{2}k_{F}} & , & \\
b=\frac{3}{32\pi ^{4}k^{2}_{F}}\left[ \left( \lambda -\Delta \right) ^{2}+\epsilon \left( \lambda -\Delta -\epsilon \right) \right]  & , & \label{23} 
\end{eqnarray}

\begin{equation}
\label{18}
U_{f}=U_{0f}-\left( cU^{2}_{0f}-2bU_{0f}^{3}\right) \ln \left( \frac{\Omega }{\omega }\right) +...
\end{equation}
with \( c=\left( \lambda -\Delta -\epsilon \right) /2\pi ^{2}k_{F} \) ,
\begin{equation}
\label{18}
U_{C}=U_{0C}-\left( dU_{0C}^{2}+2bU_{0C}^{3}\right) \ln \left( \frac{\Omega }{\omega }\right) +...
\end{equation}
with \( d=\left( 4\lambda -\left( \lambda -\Delta -\epsilon \right) /2\right) /2\pi ^{2}k_{F} \)
.

Using the \( RG \) conditions \( \omega \partial U_{0x}/\partial \omega =\omega \partial U_{0f}/\partial \omega =\omega \partial U_{0C}/\partial \omega =0 \)
the \( RG \) equations for \( U_{x} \) , \( U_{f} \) and \( U_{C} \) in
two-loop order are therefore

\begin{equation}
\label{17}
\beta \left( U_{x}\right) =\omega \frac{\partial U_{x}}{\partial \omega }=-aU_{x}^{2}+2bU_{x}^{3}+...,
\end{equation}

\begin{equation}
\label{18}
\beta \left( U_{f}\right) =\omega \frac{\partial U_{x}}{\partial \omega }=cU^{2}_{f}+2bU_{f}^{3}+...
\end{equation}
and

\begin{equation}
\label{18}
\beta \left( U_{c}\right) =\omega \frac{\partial U_{c}}{\partial \omega }=dU_{c}^{2}+2bU_{c}^{3}+...
\end{equation}

Note that there are non-trivial fixed points \( U_{x}^{*}=\frac{a}{2b} \),
\( U^{*}_{c}=-\frac{4\lambda -\left( \lambda -\Delta -\epsilon \right) /2}{2\epsilon -\left( \lambda -\Delta \right) }U_{x}^{*} \)
and \( U^{*}_{f}=-\frac{\lambda -\Delta -\epsilon }{2\epsilon -\left( \lambda -\Delta \right) }U^{*}_{x} \)
for the exchange, Cooper and forward channels respectively which are infrared
stable \( \left( IR\right)  \) but they are by no means of small magnitude
if \( k_{F}\gg \left( \lambda -\Delta \right)  \) and \( \lambda \gg \left( \lambda -\Delta \right)  \).
The magnitude of \( U_{x}^{*} \) is regulated by the ratio of \( k_{F} \)
and the size of the flat sector of \( FS \) for \( \epsilon \cong \lambda -\Delta  \).
In this case the larger the size of the flat sector with respect to \( k_{F} \)
the smaller the magnitude of \( U_{x}^{*} \). For \( U^{*}_{C} \) and \( U^{*}_{f} \)
there are extra multiplicative factors which measures basically the ratio of
widths in \( \mathbf{k} \)-space available for the divergent particle-particle
and partcle-hole diagrams in the Cooper and forward channels respectively. In
our perturbation theory scheme the expansion parameter is precisely a fraction
of \( U_{a}\left( width\right) /k_{F} \) and even a large value of the coupling
constant such as some of the \( U^{*} \)'s above presents no serious convergence
difficulty to our perturbation series expansion. 

We can use a similar \( RG \) approach for the renormalized two-particle irreducible
function with parallel spins. Now we define the corresponding one-particle irreducible
function as

\begin{equation}
\label{19}
\Gamma ^{\left( 4\right) }_{R\uparrow \uparrow }\left( p_{1},p_{2};p_{3},p_{4};U_{a};\omega \right) =\prod _{i=1}^{4}Z^{\frac{1}{2}}\left( \mathbf{p}_{i};\omega \right) \Gamma ^{\left( 4\right) }_{0\uparrow \uparrow }\left( p_{1},p_{2};p_{3},p_{4};U_{0a}\right) +A\left( \omega \right) 
\end{equation}
where \( A\left( \omega \right)  \) is an infinite additive constant since
the first term in our perturbation series expansion for \( \Gamma ^{\left( 4\right) }_{0\uparrow \uparrow } \)
is already divergent . As a result using the same choice of external momenta
as before we choose the prescription

\begin{eqnarray}
\Gamma ^{\left( 4\right) }_{R\uparrow \uparrow }\left( \mathbf{p}_{4}=\mathbf{p}_{1};\mathbf{p}_{2}=\mathbf{p}_{3};p_{0}=\omega ;U_{a};\omega \right) =Z\left( \mathbf{p}_{1};\omega \right) Z\left( \mathbf{p}_{3};\omega \right) \left[ \right.  &  & \\
\left. \Gamma ^{\left( 4\right) }_{0\uparrow \uparrow }\left( \mathbf{p}_{4}=\mathbf{p}_{1};\mathbf{p}_{2}=\mathbf{p}_{3};p_{0}=\omega ;U_{0a}\right) +A\left( \omega \right) \right] =0 &  & \label{21} 
\end{eqnarray}
Using our perturbation series result \( \left( Fig.3\right)  \) we then obtain

\begin{eqnarray}
\Gamma _{0\uparrow \uparrow }^{\left( 4\right) }\left( \mathbf{p}_{1}=\mathbf{p}_{4};\mathbf{p}_{2}=\mathbf{p}_{3},p_{0}=\omega ;U_{0a}\right) =\frac{\epsilon }{2\pi ^{2}k_{F}}U^{2}_{0f} & \ln \left( \frac{\Omega }{\omega }\right)  & \\
-\frac{U_{0f}U_{0C}^{2}}{16\pi ^{4}k_{F}^{2}}3\epsilon \left( \lambda -\Delta -\frac{\epsilon }{2}\right) \ln ^{2}\left( \frac{\Omega }{\omega }\right) +... &  & 
\end{eqnarray}
Using the same approximation \( U^{2}_{0f}\cong U^{2}_{0C}\cong U_{0}^{2} \)
as before it follows immediately

\begin{eqnarray}
A\left( \omega \right) =-\frac{\epsilon }{2\pi ^{2}k_{F}}U^{2}_{f} & \ln \left( \frac{\Omega }{\omega }\right) + & \\
+\frac{U_{f}^{3}}{16\pi ^{4}k_{F}^{2}}3\epsilon \left( \lambda -\Delta -\frac{\epsilon }{2}\right) \ln ^{2}\left( \frac{\Omega }{\omega }\right) +... &  & \label{24} 
\end{eqnarray}

Having established the existence of IR stable non-trivial fixed points in two-loop
order we can now investigate how self-energy effects produce an anomalous dimension
in the single-particle Green's function at the Fermi Surface\cite{Ferraz2}.

\section{Single-Particle Green's Function and Occupation Number at FS}

We can use the RG to calculate the renormalized Green's function \( G_{R} \)
at \( FS \). Since \( G_{R}=\left( \Gamma ^{\left( 2\right) }_{R}\right) ^{-1} \)it
follows from the previous section that

\begin{equation}
\label{18}
G_{R}\left( p_{0};\mathbf{p}^{*};\left\{ U_{a}\right\} ;\omega \right) =Z^{-1}\left( \mathbf{p}^{*};\omega \right) G_{0}\left( p_{0};\mathbf{p}^{*};\left\{ U_{0a}\right\} \right) ,
\end{equation}
where \( G_{0} \) is the corresponding bare Green's function and \( \mathbf{p}^{*} \)
is some fixed \( FS \) point. Seeing that \( G_{0} \) is independent of the
scale parameter \( \omega  \) we obtain that \( G_{R} \) satisfies the Callan-Symanzik
\( \left( CS\right)  \) equation\cite{Collins}

\begin{equation}
\label{19}
\left( \omega \frac{\partial }{\partial \omega }+\sum _{a}\beta _{a}\left( \left\{ U_{b}\right\} \right) \frac{\partial }{\partial U_{a}}+\gamma \right) G_{R}\left( p_{0};\mathbf{p}^{*};\left\{ U_{b}\right\} ;\omega \right) =0,
\end{equation}
where

\begin{equation}
\label{20}
\gamma =\omega \frac{d}{d\omega }\ln Z\left( \omega \right) 
\end{equation}
Using the fact that \( G_{R} \) at \( FS \) is a homogeneous function of only
\( \omega  \) and \( p_{0} \) of degree \( D=-1 \) it must also satisfy the
equation

\begin{equation}
\label{21}
\left( \omega \frac{\partial }{\partial \omega }+p_{0}\frac{\partial }{\partial p_{0}}\right) G_{R}\left( p_{0};\mathbf{p}^{*};\left\{ U_{a}\right\} ;\omega \right) =-G_{R}\left( p_{0};\mathbf{p}^{*};\left\{ U_{a}\right\} ;\omega \right) 
\end{equation}
Combining this with the \( CS \) equation we then find

\begin{equation}
\label{22}
\left( -p_{0}\frac{\partial }{\partial p_{0}}+\sum _{a}\beta _{a}\left( \left\{ U_{b}\right\} \right) \frac{\partial }{\partial U_{a}}+\gamma -1\right) G_{R}\left( p_{0};\mathbf{p}^{*};\left\{ U_{b}\right\} ;\omega \right) =0
\end{equation}
However up to two-loop order we don't need to distinguish the mixing effects
of the different scattering channels in the self-energy and for simplicity we
can assume that the divergences up to this order are entirely due to the exchange
channel. Thus the \( CS \) equation reduces to

\begin{equation}
\label{23}
\left( -p_{0}\frac{\partial }{\partial p_{0}}+\beta _{x}\left( U_{x}\right) \frac{\partial }{\partial U_{x}}+\gamma -1\right) G_{R}\left( p_{0};\mathbf{p}^{*};U_{x};\omega \right) =0
\end{equation}
From this we obtain that the formal solution for \( G_{R} \) is

\begin{equation}
\label{23}
G_{R}\left( p_{0};\mathbf{p}^{*};U_{x};\omega \right) =\frac{1}{p_{0}}\exp \left( \int ^{p_{0}}_{\omega }d\ln \left( \frac{\overline{p}_{0}}{\omega }\right) \gamma \left[ U_{x}\left( \overline{p}_{0};\mathbf{p}^{*};U_{x}\right) \right] \right) ,
\end{equation}
where

\begin{equation}
\label{24}
\frac{dU_{x}\left( \overline{p}_{0};U\right) }{d\ln \left( \frac{\overline{p}_{0}}{\omega }\right) }=\beta _{x}\left( U_{x}\left( \overline{p}_{0};\mathbf{p}^{*};U_{x}\right) \right) ,
\end{equation}
with \( U_{x}\left( \overline{p}_{0}=\omega ;\mathbf{p}^{*};U_{x}\right) =U_{x}\left( \mathbf{p}^{*};\omega \right)  \).

If we assume that as the physical system approaches the Fermi Surface as \( p_{0}\sim \omega \rightarrow 0 \),
it also acquires a critical condition with the running coupling constant \( U_{x}\left( \mathbf{p}^{*};\omega \right) \rightarrow U_{x}^{*}\left( \mathbf{p}^{*}\right)  \)
and if we take \( \mathbf{p}^{*}=\left( \lambda -\epsilon ,-k_{F}+\frac{\Delta ^{2}}{2k_{F}}\right)  \)
we can use our perturbation theory result for \( Z\left( \mathbf{p}^{*};\omega \right)  \)
up to order \( O\left( U_{x}^{*2}\right)  \) to obtain

\begin{equation}
\label{25}
\gamma =\frac{3U_{x}^{*2}}{16\pi ^{4}k^{2}_{F}}\left[ \frac{\left( \lambda -\Delta \right) ^{2}}{2}+\epsilon \left( \lambda -\Delta -\epsilon \right) \right] +...=\gamma ^{*}
\end{equation}
As a result of this \( G_{R} \) develops an anomalous dimension given by\cite{Ferraz2}

\begin{equation}
\label{26}
G_{R}\left( p_{0};\mathbf{p}^{*};U_{x}^{*};\omega \right) =\frac{1}{\omega }\left( \frac{\omega ^{2}}{p_{0}^{2}}\right) ^{\frac{1-\gamma ^{*}}{2}},
\end{equation}
 If we make the analytical continuation \( p_{0}\rightarrow p_{0}+i\delta  \)
, \( G_{R} \) reduces to at \( FS \)

\begin{equation}
\label{27}
G_{R}\left( p_{0};U^{*};\omega \right) =-\frac{1}{\omega }\left( \frac{\omega ^{2}}{p_{0}^{2}}\right) ^{\left( \frac{1-\gamma ^{*}}{2}\right) }\left[ \cos \left( \pi \gamma ^{*}\right) +i\sin \left( \pi \gamma ^{*}\right) \right] 
\end{equation}
Using this result the spectral function \( A\left( k_{F},p_{0}\right) =-ImG_{R} \)
becomes

\begin{equation}
\label{28}
A\left( k_{F},p_{0}\right) =\left| \frac{p_{0}}{\omega }\right| ^{\gamma ^{*}}\frac{\sin \left( \pi \gamma ^{*}\right) }{\left| p_{0}\right| },
\end{equation}
and the number density \( n\left( k_{F}\right)  \) reduces to

\begin{equation}
\label{29}
n\left( k_{F}\right) =\frac{1}{2}\frac{\sin \left( \pi \gamma ^{*}\right) }{\pi \gamma ^{*}}
\end{equation}
Notice that if \( U_{x}^{*}\rightarrow 0 \), \( \gamma ^{*}\rightarrow 0 \)
and as a result \( n\left( k_{F};U^{*}_{x}=0\right) =\frac{1}{2} \). Alternatively
if we replace our two-loop value for \( U_{x}^{*} \) we get

\begin{equation}
\label{30}
\gamma ^{*}=\frac{4}{3}\left( 2\epsilon -\left( \lambda -\Delta \right) \right) ^{2}\frac{\left[ \left( \lambda -\Delta \right) ^{2}/2+\epsilon \left( \lambda -\Delta -\epsilon \right) \right] }{\left[ \left( \lambda -\Delta \right) ^{2}+\epsilon \left( \lambda -\Delta -\epsilon \right) \right] ^{2}}
\end{equation}
If we now take \( \epsilon =\frac{2}{3}\left( \lambda -\Delta \right)  \) we
find \( \gamma ^{*}\cong .07 \),

\begin{equation}
\label{28}
ImG_{R}\left( p_{0};U^{*};\omega \right) \cong -\left( \frac{\omega ^{2}}{p_{0}^{2}}\right) ^{-.035}\frac{1}{\left| p_{0}\right| },
\end{equation}
and as a result \( n\left( k_{F};U^{*}_{x}\right) \cong .14 \). This result
shows that there is indeed no discontinuity at \( n\left( k_{F}\right)  \).
Moreover there is only a small correction to the marginal Fermi liquid result
for the 'cold' spot point which suffers the direct effect of the flat sectors
through \( \Sigma  \). The correction to the linear behavior of \( Im\Sigma  \)
is practically not observed experimentally. The power law behavior of \( G_{R} \)
and the value of \( n\left( k_{F}\right)  \) independent of the sign of the
coupling constant resembles the results obtained for a Luttinger liquid\cite{Mattis}.
However for the one-dimensional Luttinger liquid \( \partial n\left( p\right) /\partial p\mid _{p=k_{F}}\rightarrow \infty  \).
In order to see if the occupation function shows the same behavior in our case
we have to generalize our \( CS \) equation to explicitly include the momentum
dependence for \( G_{R} \) in the vicinity of a given 'cold' spot point.

\section{Green's Function and Momentum Distribution Function near a 'Cold' Spot Point}

Let us choose for simplicity the point \( \mathbf{p}=\left( \Delta -\upsilon ,-k_{F}+\frac{\Delta ^{2}}{2k_{F}}-\frac{\upsilon \Delta }{k_{F}}\right)  \)
in the \( \left( 0,-k_{F}\right)  \)-patch of our FS model. Quite generically
the relation between the renormalized and bare one-particle irreducible \( \Gamma ^{\left( 2\right) '}s \)
holds for any momentum value. Thus taking into consideration our perturbative
two-loop self-energy result together with the fact that at \( p_{0}=0 \) and
in the vicinity of the \( FS \) point \( \left( \Delta -\upsilon ,-k_{F}+\frac{\Delta ^{2}}{2k_{F}}-\frac{\upsilon \Delta }{k_{F}}\right)  \),
it is natural to define a renormalized \( \left( k_{F}\right) _{R} \) such
that \( \Gamma ^{\left( 2\right) }_{R} \) reduces to

\begin{eqnarray}
\Gamma ^{\left( 2\right) }_{R}\left( p_{0}=0,\mathbf{p};\omega \right) =\overline{p}=\left( k_{F}\left( p_{y}+k_{F}-\frac{\Delta ^{2}}{2k_{F}}+\frac{\upsilon \Delta }{k_{F}}\right) \right)  &  & \\
=Z\left( \omega \right) k_{F}\left( p_{y}+k_{F}-\frac{\Delta ^{2}}{2k_{F}}+\frac{\upsilon \Delta }{k_{F}}\right)  &  & \label{29} 
\end{eqnarray}
In the presence of a non-zero \( \overline{p} \) the \( CS \) equation for
\( G_{R} \) in the neighborhood of this 'cold' spot point becomes

\begin{equation}
\label{30}
\left( \omega \frac{\partial }{\partial \omega }+\beta \left( U_{x}\right) \frac{\partial }{\partial U_{x}}+\gamma \overline{p}\frac{\partial }{\partial \overline{p}}+\gamma \right) G_{R}\left( p_{0};\overline{p};U_{x};\omega \right) =0
\end{equation}
Since now we have that

\begin{equation}
\label{31}
\left( \omega \frac{\partial }{\partial \omega }+p_{0}\frac{\partial }{\partial p_{0}}+\overline{p}\frac{\partial }{\partial \overline{p}}\right) G_{R}\left( p_{0};\overline{p};U_{x};\omega \right) =-G_{R}\left( p_{0};\overline{p};U_{x};\omega \right) 
\end{equation}
it follows from this that \( G_{R} \) satisfies the \( RG \) equation

\begin{equation}
\label{32}
\left( p_{0}\frac{\partial }{\partial p_{0}}+\left( 1-\gamma \right) \overline{p}\frac{\partial }{\partial \overline{p}}-\beta \left( U_{x}\right) \frac{\partial }{\partial U_{x}}+1-\gamma \right) G_{R}\left( p_{0};\overline{p};U_{x};\omega \right) =0.
\end{equation}
We can therefore write \( G_{R} \) in the form

\begin{equation}
\label{33}
G_{R}\left( p_{0};\overline{p};U_{x};\omega \right) =\mathcal{G}\left( U_{x}\left( p_{0};U_{x}\right) ;\overline{p}\left( p_{0};\overline{p}\right) \right) \exp -\int _{\omega }^{p_{0}}d\ln \left( \frac{\overline{p}_{0}}{\omega }\right) \left[ 1-\gamma \left( U_{x}\left( \overline{p}_{0};\overline{p};U_{x}\right) \right) \right] ,
\end{equation}
where

\begin{equation}
\label{34}
\overline{p}\left( p_{0};U_{x}\right) =\overline{p}\exp \left( -\int ^{p_{0}}_{\omega }d\ln \left( \frac{\overline{p}_{0}}{\omega }\right) \left[ 1-\gamma \left( U_{x}\left( \overline{p}_{0};\overline{p};U_{x}\right) \right) \right] \right) ,
\end{equation}
and the \( \beta  \) -function is determined perturbatively. If we assume as
before that the physical system is brought to criticality as \( \omega \rightarrow 0 \)
and \( U_{x}\left( \omega \right) \rightarrow U_{x}^{*}\neq 0 \) we can use
our perturbation theory result for \( \gamma  \) and these equations reduce
to

\begin{equation}
\label{35}
G_{R}\left( p_{0};\overline{p};U_{x}^{*};\omega \right) =\frac{1}{p_{0}}\mathcal{G}\left( \overline{p}\left( p_{0};U_{x}^{*}\right) \right) \left( \frac{p_{0}}{\omega }\right) ^{\gamma ^{*}},
\end{equation}
with

\begin{equation}
\label{36}
\overline{p}\left( p_{0};U^{*}_{x}\right) =\overline{p}\left( \frac{p_{0}}{\omega }\right) ^{\left( \gamma ^{*}-1\right) }.
\end{equation}
The function \( \mathcal{G} \) is determined from perturbation theory. Recalling
that at zeroth-order, for \( p_{0}\simeq \omega  \), we have that 
\begin{equation}
\label{37}
G_{R}\cong \frac{1}{\omega +\overline{p}}+O\left( U_{x}^{*2}\right) ,
\end{equation}
and it turns ou that

\begin{equation}
\label{38}
\mathcal{G}\left( \overline{p}\left( p_{0};U_{x}^{*}\right) \right) =\frac{\omega }{\omega +\overline{p}\left( p_{0}:U^{*}\right) }+...
\end{equation}
Finally, combining all these results we get that in the vicinity of our 'cold'
spot point

\begin{equation}
\label{39}
G_{R}\left( p_{0};\overline{p};U_{x}^{*};\omega \right) =\frac{1}{p_{0}}\left( \frac{p^{2}_{0}}{\omega ^{2}}\right) ^{\frac{\gamma ^{*}}{2}}\left[ 1+\frac{\overline{p}}{p_{0}}\left( \frac{p_{0}^{2}}{\omega ^{2}}\right) ^{\frac{\gamma ^{*}}{2}}\right] ^{-1}
\end{equation}
If we now do again the analytic continuation making \( p_{0}\rightarrow p_{0}+i\delta  \)
we obtain the renormalized Green's function as

\begin{equation}
\label{40}
G_{R}\left( p_{0};\overline{p};U_{x}^{*}\right) =\frac{\left( \frac{p^{2}_{0}}{\omega ^{2}}\right) ^{\frac{\gamma ^{*}}{2}}}{\overline{p}\left( \frac{p^{2}_{0}}{\omega ^{2}}\right) ^{^{\frac{\gamma ^{*}}{2}}}-\left| p_{0}\right| \cos \left( \pi \gamma ^{*}\right) +i\left| p_{0}\right| \sin \left( \pi \gamma ^{*}\right) }
\end{equation}
It follows from this that imaginary part of the renormalized self-energy \( Im\Sigma _{R} \)
is given by

\begin{equation}
\label{41}
Im\Sigma _{R}\left( p_{0};\overline{p};U_{x}^{*};\omega \right) =-\left| p_{0}\right| \left( \frac{p^{2}_{0}}{\omega ^{2}}\right) ^{-\frac{\gamma ^{*}}{2}}\sin \left( \pi \gamma ^{*}\right) ,
\end{equation}
and the renormalized spectral function \( A_{R}\left( \overline{p};\omega \right)  \)
becomes\cite{Ferraz2}

\begin{equation}
\label{42}
A_{R}\left( \overline{p};\omega \right) =\frac{\left| p_{0}\right| \left( \frac{p^{2}_{0}}{\omega ^{2}}\right) ^{-\frac{\gamma ^{*}}{2}}\sin \left( \pi \gamma ^{*}\right) }{\left( \overline{p}-\left| p_{0}\right| \left( \frac{\omega ^{2}}{p_{0}^{2}}\right) ^{\frac{\gamma ^{*}}{2}}\cos \left( \pi \gamma ^{*}\right) \right) ^{2}+p^{2}_{0}\left( \frac{\omega ^{2}}{p^{2}_{0}}\right) ^{\gamma ^{*}}\sin ^{2}\left( \pi \gamma ^{*}\right) }
\end{equation}
We can immediately infer from this result that our renormalized Fermi Surface
near the given 'cold' spot point is now characterized by a dispersion law given
by

\begin{equation}
\label{42}
p_{0}=\varepsilon \left( \overline{p}\right) =\pm \left( \frac{\left| \overline{p}\right| \sec \left( \pi \gamma ^{*}\right) }{\omega ^{\gamma ^{*}}}\right) ^{\frac{1}{1-\gamma ^{*}}},
\end{equation}
which in turn produces a ``Fermi velocity'' \( v_{F} \) given by

\begin{equation}
\label{43}
v_{F}=\frac{k_{F}}{1-\gamma ^{*}}\left( \sec \left( \pi \gamma ^{*}\right) \right) ^{\frac{1}{1-\gamma ^{*}}}\left( \frac{\left| \overline{p}\right| }{\omega }\right) ^{\frac{\gamma ^{*}}{1-\gamma ^{*}}}
\end{equation}
Clearly for \( \gamma ^{*}/\left( 1-\gamma ^{*}\right) >0 \), we have that
\( v_{F}\rightarrow 0 \) if \( \left| \overline{p}\right| /\omega \rightarrow 0 \)
and is finite for \( \frac{\overline{p}}{\omega }=1 \). 

Finally, using our spectral function result we can calculate the momentum distribution
function \( n\left( \overline{p}\right)  \) for \( \frac{\overline{p}}{\omega }\sim 0 \).
We obtain for non- integers \( \gamma ^{*}/\left( 1-\gamma ^{*}\right)  \)
and \( \left( 2-\gamma ^{*}\right) /\left( 1-\gamma ^{*}\right)  \)

\begin{eqnarray}
n\left( \overline{p}\right) \cong \frac{\left| \sin \left( \pi \gamma ^{*}\right) \right| }{2\pi \gamma ^{*}}\left\{ \left[ 1+\frac{\gamma ^{*}}{2\gamma ^{*}-1}\frac{2\overline{p}}{\omega }\cos \left( \pi \gamma ^{*}\right) +...\right] \right.  &  & \\
+\frac{\gamma ^{*}}{1-\gamma ^{*}}\cos \left( \frac{\pi }{1-\gamma ^{*}}\right) \Gamma \left( \frac{2-\gamma ^{*}}{1-\gamma ^{*}}\right) \Gamma \left( \frac{-\gamma ^{*}}{1-\gamma ^{*}}\right)  &  & \\
\left. \left[ \frac{2\overline{p}}{\omega }\cos \left( \pi \gamma ^{*}\right) \right] ^{\frac{\gamma ^{*}}{1-\gamma ^{*}}}\right\}  &  & \label{44} 
\end{eqnarray}

It follows from this that

\begin{equation}
\label{45}
\frac{\partial n\left( \overline{p}\right) }{\partial \overline{p}}\sim \left( \frac{\left| \overline{p}\right| }{\omega }\right) ^{-\left( \frac{1-2\gamma ^{*}}{1-\gamma ^{*}}\right) }
\end{equation}
Therefore if \( 1>2\gamma ^{*} \) or \( \gamma ^{*}>1 \) we have \( \partial n\left( \overline{p}\right) /\partial \overline{p}\rightarrow \infty  \)
when \( \overline{p}/\omega \rightarrow 0 \). If we use our two-loop perturbation
scheme and assume that \( \upsilon \ll \epsilon =\frac{2}{3}\left( \lambda -\Delta \right)  \)
we can use the value of \( U^{*}_{x} \) obtained before. Combining this with
our perturbative result for \( Z\left( \omega \right)  \) which at the appropriate
momentum value is given by

\begin{eqnarray}
Z\left( \omega \right) =1-\frac{3}{32\pi ^{4}k^{2}_{F}}\left( \lambda -\Delta -\upsilon \right) ^{2}U^{*2}_{x}\ln \left( \frac{\Omega }{\omega }\right) +... &  & \\
\cong 1-\frac{3}{32\pi ^{4}k^{2}_{F}}\left( \lambda -\Delta \right) ^{2}U^{*2}_{x}\ln \left( \frac{\Omega }{\omega }\right)  & \label{46} 
\end{eqnarray}
we find \( \gamma ^{*}\cong 6/121 \). For this value of \( \gamma ^{*} \)the
momentum distribution function is clearly non-analytic at \( \overline{p}=0 \)
indicating that some remains of a Fermi surface continues to be present in the
system. Thus for this \( \mathbf{k}- \)space region the physical system resembles
indeed a Luttinger liquid\cite{Mattis}. Moreover despite the fact that both
\( \partial Re\Sigma /\partial \omega  \) and \( \partial Re\Sigma /\partial k \)
are both singular at \( FS \) the renormalized Fermi velocity \( v_{F} \)
can continue to be finite as in a incompressible fluid\cite{Anderson} if \( \overline{p}/\omega =1 \).
This shows that the effects produced by the flat sectors of \( FS \) leads
to a complete breakdown of the Landau quasiparticle picture in the ``cold''
spots. This is in general agreement with recent photoemission data\cite{Valla}
for optimally doped Bi2212 which report a marginal Fermi liquid behavior for
\( Im\Sigma  \) and a large broadening of the spectral peak even around the
\( \left( \frac{\pi }{2},\frac{\pi }{2}\right)  \) region of the Fermi Surface
for temperatures higher than \( T_{c} \). However our results depend in an
important way on the value of the non-trivial fixed point \( U^{*} \). It is
therefore opportune to check what happens to our results if we include higher-order
corrections. To estimate this we discuss the higher-loop contributions to both
the quasiparticle \( \Sigma _{\uparrow } \) and \( \Gamma ^{\left( 4\right) }_{\uparrow \downarrow ;\uparrow \downarrow } \).

\section{Higher-Order Corrections}

In 3-loop order with our local subtraction regularization method we don't distinguish
the different bare coupling functions at order \( O\left( U^{3}_{0}\right)  \).
There are in this way two contributions to the bare self-energy \( \Sigma _{0\uparrow } \)
\( \left( Fig.4\right)  \):

\begin{eqnarray}
\Sigma ^{\left( a\right) }_{0\uparrow }\left( p\right) =U_{0x}^{3}\int _{q}G^{\left( 0\right) }_{\downarrow }\left( q\right) \left( \chi _{\uparrow \downarrow }^{\left( 0\right) }\left( q-p\right) \right) ^{2} &  & \\
=U_{0C}U_{0x}^{2}\int _{q}G^{\left( 0\right) }_{\downarrow }\left( q\right) \left( \chi _{\uparrow \downarrow }^{\left( 0\right) }\left( q-p\right) \right) ^{2} &  & \\
=U_{0}^{3}\int _{q}G^{\left( 0\right) }_{\downarrow }\left( q\right) \left( \chi _{\uparrow \downarrow }^{\left( 0\right) }\left( q-p\right) \right) ^{2} &  & \label{65} 
\end{eqnarray}
and

\begin{eqnarray}
\Sigma ^{\left( b\right) }_{0\uparrow }\left( p\right) =U_{0C}^{3}\int _{q}G^{\left( 0\right) }_{\downarrow }\left( q\right) \left( \Pi _{\uparrow \downarrow }^{\left( 0\right) }\left( q-p\right) \right) ^{2} &  & \\
=U_{0x}U_{0C}^{2}\int _{q}G^{\left( 0\right) }_{\downarrow }\left( q\right) \left( \Pi _{\uparrow \downarrow }^{\left( 0\right) }\left( q-p\right) \right) ^{2} &  & \label{66} \\
=U_{0}^{3}\int _{q}G^{\left( 0\right) }_{\downarrow }\left( q\right) \left( \Pi _{\uparrow \downarrow }^{\left( 0\right) }\left( q-p\right) \right) ^{2} &  & 
\end{eqnarray}
with \( U_{0C}^{3}\cong U_{0x}^{3}\cong U_{0}^{3} \). However since for \( \mathbf{p}=\left( \Delta ,-k_{F}+\frac{\Delta ^{2}}{2k_{F}}\right)  \)
we have that \( \Pi ^{\left( 0\right) }\left( \mathbf{q}+\mathbf{p};q_{0}+p_{0}\right) =-\chi ^{\left( 0\right) }_{\uparrow \downarrow }\left( \mathbf{q}-\mathbf{p};q_{0}-\left( -p_{0}\right) \right)  \)
these two contributions at this order cancel each other exactly for \( p_{0}=\omega \cong 0 \).
The next non-zero contributions are therefore produced by the fourth-order terms
\( \left( Fig.5\right)  \). They all have the same relative sign bringing about
a strong mix between the different scattering channels. Their calculation is
non-trivial and it is beyond the scope of this present work. However if we include
three-loop contributions for the various renormalized couplings we can distinguish
the different bare couplings functions at order \( O\left( U_{0}^{2}\right)  \)
and this produces important changes in our results. To observe this in detail
we consider the quasiparticle weight \( Z\left( \mathbf{p},\omega \right)  \)
for \( \mathbf{p}=\left( \Delta ,-k_{F}+\frac{\Delta ^{2}}{2k_{F}}\right)  \).
We find now

\begin{equation}
\label{21}
Z^{-1}\left( \mathbf{p},\omega \right) =1+\frac{3}{64\pi ^{4}k^{2}_{F}}\left( \lambda -\Delta \right) ^{2}\ln \left( \frac{\Omega }{\omega }\right) \left( U^{2}_{0x}+U^{2}_{0C}\right) +...
\end{equation}
 A similar result also applies for \( Z\left( \mathbf{p}^{*},\omega \right)  \)
for \( \mathbf{p}^{*}=\left( \lambda -\epsilon ,-k_{F}+\frac{\Delta ^{2}}{2k_{F}}\right)  \).
If we repeat the same procedure as before but now distinguishing the diverse
bare coupling functions at two-loop order we find respectively

\begin{eqnarray}
U_{x}\left( \mathbf{p}_{1}-\mathbf{p}_{4};\omega \right) =U_{0x}+\left[ \frac{\epsilon }{2\pi ^{2}k_{F}}U^{2}_{0x}+\right.  &  & \\
\left. -bU_{0x}\left( U^{2}_{0x}+U^{2}_{0C}\right) +\right.  &  & \\
\left. -\frac{\left( \lambda -\Delta -\epsilon \right) }{2\pi ^{2}k_{F}}U^{2}_{0C}\right] \ln \left( \frac{\Omega }{\omega }\right) +..., & , & \label{22} 
\end{eqnarray}

\begin{eqnarray}
U_{f}\left( \mathbf{p}_{1}-\mathbf{p}_{3};\omega \right) =U_{0f}-\left[ \frac{\left( \lambda -\Delta -\epsilon \right) }{2\pi ^{2}k_{F}}U^{2}_{0C}+\right.  &  & \\
\left. +bU_{0f}\left( U^{2}_{0x}+U^{2}_{0C}\right) \right] \ln \left( \frac{\Omega }{\omega }\right) +..., & , & \label{23} 
\end{eqnarray}

\begin{eqnarray}
 & U_{C}\left( \mathbf{p}_{1}=-\mathbf{p}_{2};\omega \right) =U_{0C}-\left[ \frac{4\lambda }{2\pi ^{2}k_{F}}U^{2}_{0C}+\right.  & \\
 & \left. +bU_{0x}\left( U^{2}_{0x}+U^{2}_{0C}\right) +\right.  & \\
 & \left. -\frac{\left( \lambda -\Delta -\epsilon \right) }{4\pi ^{2}k_{F}}U^{2}_{0C}\right] \ln \left( \frac{\Omega }{\omega }\right)  & \label{24} 
\end{eqnarray}
for the exchange, forward and Cooper channels. If we now define the corresponding
\( \beta  \)-functions as

\begin{eqnarray}
\beta _{x}\left( U_{x},U_{f},U_{C}\right) =\omega \frac{\partial U_{x}}{\partial \omega } &  & \\
\beta _{f}\left( U_{f},U_{x},U_{C}\right) =\omega \frac{\partial U_{f}}{\partial \omega } &  & \\
\beta _{C}\left( U_{C},U_{x},U_{f}\right) =\omega \frac{\partial U_{C}}{\partial \omega } &  & \label{25} 
\end{eqnarray}
it follows that

\begin{eqnarray}
\beta _{x}\left( U_{x},U_{f},U_{C}\right) =-\frac{\epsilon }{2\pi ^{2}k_{F}}U^{2}_{x}+bU_{x}\left( U^{2}_{x}+U^{2}_{C}\right) + &  & \\
+\frac{\left( \lambda -\Delta -\epsilon \right) }{2\pi ^{2}k_{F}}U^{2}_{C}+..., & , & \label{26} 
\end{eqnarray}

\begin{equation}
\label{27}
\beta _{f}\left( U_{f},U_{x},U_{C}\right) =\frac{\left( \lambda -\Delta -\epsilon \right) }{2\pi ^{2}k_{F}}U^{2}_{C}+bU_{f}\left( U^{2}_{x}+U^{2}_{C}\right) +...
\end{equation}
and

\begin{eqnarray}
\beta _{C}\left( U_{C},U_{x},U_{f}\right) =\frac{4\lambda }{2\pi ^{2}k_{F}}U^{2}_{C} & +bU_{C}\left( U^{2}_{x}+U^{2}_{C}\right)  & \\
- & \frac{\left( \lambda -\Delta -\epsilon \right) }{4\pi ^{2}k_{F}}U^{2}_{x}+... & \label{28} 
\end{eqnarray}
with \( b=\gamma ^{*}/2 \) given as before. To determine the fixed points let
us choose for simplicity the case \( \epsilon =\lambda -\Delta  \). Taking
\( \beta _{x}=\beta _{f}=\beta _{C}=0 \) it follows immediately that the non-trivial
fixed points for this value of \( \epsilon  \) are

\begin{eqnarray}
U_{x}^{*}=\frac{16\pi ^{2}k_{F}}{3\left( \lambda -\Delta \right) }\zeta ^{-1}, &  & \\
U_{f}^{*}=0, &  & \\
U_{C}^{*}=-\frac{4\pi ^{2}k_{F}}{3\lambda }\zeta ^{-1} &  & \label{29} 
\end{eqnarray}
with \( \zeta =\left[ 1+\left( \frac{\lambda -\Delta }{4\lambda }\right) ^{2}\right] >1 \).
Defining the matrix of eigenvalues \( M_{ij} \) by

\begin{equation}
\label{30}
M_{ij}=\left( \frac{\partial \beta _{i}}{\partial U_{j}}\right) _{U^{*}}
\end{equation}
for \( i,j=C,x,f \) respectively we can expand in coupling space around these
fixed points to find\cite{Weinberg}

\begin{equation}
\label{31}
\beta _{i}\cong \sum _{j}M_{ij}\left( U_{j}\left( \mathbf{p}^{*};\omega \right) -U_{j}^{*}\left( \mathbf{p}^{*}\right) \right) +...,
\end{equation}
Integrating these out we obtain

\begin{equation}
\label{32}
U_{i}=U^{*}_{i}+\sum _{j}c_{j}V^{i}_{j}\omega ^{\gamma _{j}}
\end{equation}
where

\begin{equation}
\label{33}
M_{ij}V_{ij}=\gamma _{i}V_{i}
\end{equation}
Using our results it then turns out that

\begin{equation}
\label{32}
U_{C}\left( \mathbf{p}^{*};\omega \right) \cong U^{*}_{C}-\frac{c_{1}}{\sqrt{2\zeta }}\left( \frac{\lambda -\Delta }{4\lambda }\right) \omega ^{\frac{8}{3}\zeta ^{-1}}+\frac{c_{2}}{\sqrt{\zeta }}\omega ^{-\frac{8}{3}\zeta ^{-1}},
\end{equation}

\begin{equation}
\label{33}
U_{x}\left( \mathbf{p}^{*};\omega \right) \cong U_{x}^{*}+\frac{c_{1}}{\sqrt{2\zeta }}\omega ^{\frac{8}{3}\zeta ^{-1}}+\frac{c_{2}}{\sqrt{\zeta }}\left( \frac{\lambda -\Delta }{4\lambda }\right) \omega ^{-\frac{8}{3}\zeta ^{-1}},
\end{equation}
and

\begin{equation}
\label{34}
U_{f}\left( \mathbf{p}^{*};\omega \right) \cong \frac{c_{1}}{\sqrt{2}}\omega ^{\frac{8}{3}\zeta ^{-1}}
\end{equation}
where \( c_{1} \), and \( c_{2} \) are constants. As a result unless there
is one adjustable parameter which can be tunned to produce \( c_{2}=0 \) we
no longer approach the fixed point \( \left( U_{C}^{*},U^{*}_{x},U_{f}^{*}\right)  \)
as we approach the Fermi Surface when taking the limit \( \omega \rightarrow 0 \).
This is the main effect produced by the mixing of scattering channels at higher
order perturbation theory. Now the running coupling functions \( U_{C}\left( \mathbf{p};\omega \right)  \)
and \( U\left( \mathbf{p};\omega \right)  \)are only infrared stable if there
exists an external parameter which could be, for example, either temperature
or hole concentration which can be readily adjusted to nullify \( c_{2} \)
at \( FS \). The critical surface formed by the set of trajectories of \( U_{i}\left( \mathbf{p}^{*};\omega \right)  \)
which are attracted into the fixed point \( \left( U_{C}^{*},U_{x}^{*},U_{f}^{*}\right)  \)
for \( \omega \rightarrow 0 \) has in this way codimensionality one. If we
assume that the external parameter needed is the physical quantity \( \theta  \),
in the vicinity of the phase transition the coupling constants associated with
the three scattering channels become

\begin{equation}
\label{35}
U_{C}\left( \omega \right) \cong U^{*}_{C}+-\frac{c_{1}}{\sqrt{2\zeta }}\left( \frac{\lambda -\Delta }{4\lambda }\right) _{1}\omega ^{\frac{8}{3}\zeta ^{-1}}+\frac{\left( \theta -\theta _{c}\right) }{\sqrt{\zeta }}\omega ^{-\frac{8}{3}\zeta ^{-1}},
\end{equation}

\begin{equation}
\label{36}
U_{x}\left( \omega \right) \cong U_{x}^{*}+\frac{c_{1}}{\sqrt{2\zeta }}\omega ^{\frac{8}{3}\zeta ^{-1}}+\frac{\left( \theta -\theta _{c}\right) }{\sqrt{\zeta }}\left( \frac{\lambda -\Delta }{4\lambda }\right) \omega ^{-\frac{8}{3}\zeta ^{-1}},
\end{equation}
and

\begin{equation}
\label{37}
U_{f}\left( \omega \right) \cong \frac{c_{1}}{\sqrt{2}}\omega ^{\frac{8}{3}\zeta ^{-1}}
\end{equation}
where \( \theta _{c} \) is the critical value of \( \theta  \) at the transition
point.

\section{Spin and Charge Susceptibilities}

Let us consider initially the longitudinal spin ( charge )susceptibility \( \chi _{zz}\left( q\right)  \)
\( \left( \chi _{c}\left( q\right) \right)  \) for \( \left| \mathbf{q}\right| =\left| \mathbf{q}^{*}\right| \cong 2k_{F}-\frac{\Delta ^{2}}{k_{F}}+\frac{\Delta \delta }{k_{F}} \)
which as expected is singular in the exchange ( forward )channel. For consistency
we define

\begin{equation}
\label{28}
\chi _{zz\left( c\right) }\left( q\right) =\Gamma ^{\left( 0,2\right) }_{zz\left( c\right) }\left( k\right) =\frac{\delta ^{2}}{\delta h\left( q\right) \delta h\left( -q\right) }\left\langle \exp -\left( \int _{k}h\left( k\right) \xi \left( k\right) \right) \right\rangle _{h=0},
\end{equation}
where

\begin{equation}
\label{29}
\xi _{s}\left( k\right) =S_{z}\left( k\right) =\int _{p}\left( \psi _{\uparrow }^{\dagger }\left( p+k\right) \psi _{\uparrow }\left( p\right) -\psi ^{\dagger }_{\downarrow }\left( p+k\right) \psi _{\downarrow }\left( p\right) \right) ,
\end{equation}
for the longitudinal spin susceptibility, or

\begin{equation}
\label{30}
\xi _{c}\left( k\right) =\int _{p}\left( \psi _{\uparrow }^{\dagger }\left( p+k\right) \psi _{\uparrow }\left( p\right) +\psi ^{\dagger }_{\downarrow }\left( p+k\right) \psi _{\downarrow }\left( p\right) \right) ,
\end{equation}
for the corresponding charge susceptibility, with

\begin{equation}
\label{31}
\left\langle ...\right\rangle =\int d\left[ \psi ^{\dagger }_{\sigma }\right] d\left[ \psi _{\sigma }\right] \exp -\int \mathcal{L}_{E}\left[ \psi ^{\dagger }_{\sigma };\psi _{\sigma }\right] .
\end{equation}

Here \( \mathcal{L}_{E} \) is the Euclidean version of the single-particle
lagrangian given by Eq.\( \left[ 1\right]  \). Using perturbation theory up
to one-loop level with yet no mixing of scattering channels and with the bare
coupling as the expansion parameter we obtain to order \( O\left( U_{0}\right)  \)
that \( \left( Fig.6\right)  \)

\begin{equation}
\label{31}
\Gamma ^{\left( 0,2\right) }_{0zz\left( c\right) }\left( q_{0};\mathbf{q}^{*};U_{0}\right) =2\chi ^{\left( 0\right) }\left( q\right) \pm 2U_{0}\left( \chi ^{\left( 0\right) }\left( q\right) \right) ^{2}\pm ...
\end{equation}
Since \( \chi ^{\left( 0\right) }\left( q_{0},\mathbf{q}^{*}\right)  \) is
already logarithmic divergent to define a finite renormalized \( \Gamma ^{\left( 0,2\right) }_{Rzz} \)
we have to introduce a new composite field scale multiplicative factor \( Z_{\chi _{s}} \)
and a constant term. We have that

\begin{equation}
\label{32}
\Gamma ^{\left( 0,2\right) }_{Rzz}\left( q_{0},\mathbf{q}^{*}:U;\omega \right) =Z_{\chi _{s}}^{2}\left( \mathbf{q}^{*};\omega \right) \Gamma ^{\left( 0,2\right) }_{0zz}\left( q_{0},\mathbf{q}^{*};U_{0}\right) +C\left( \omega \right) 
\end{equation}
or

\begin{equation}
\label{33}
\Gamma ^{\left( 0,2\right) }_{Rc}\left( q_{0},\mathbf{q}^{*}:U;\omega \right) =Z_{\chi _{c}}^{2}\left( \mathbf{q}^{*};\omega \right) \Gamma ^{\left( 0,2\right) }_{0c}\left( q_{0},\mathbf{q}^{*};U_{0}\right) +D\left( \omega \right) 
\end{equation}

The constants \( C \) and \( D \) above can be infinite but they disappear
from the problem when we differentiate equation \( \left( 100\right)  \) or
\( \left( 101\right)  \) with respect to \( \ln q_{0} \) \cite{Solyom}. We
get in this case

\begin{equation}
\label{33}
\frac{\partial }{\partial \ln q_{0}}\Gamma ^{\left( 0,2\right) }_{Rzz\left( c\right) }\left( q_{0},\mathbf{q}^{*}:U;\omega \right) =Z_{\chi _{s\left( c\right) }}^{2}\left( \mathbf{q}^{*};\omega \right) \frac{\partial }{\partial \ln q_{0}}\Gamma ^{\left( 0,2\right) }_{0zz\left( c\right) }\left( q_{0},\mathbf{q}^{*};U_{0}\right) 
\end{equation}
It follows from this that \( \Phi ^{\left( 0,2\right) }_{Rzz\left( c\right) }\left( q_{0};\mathbf{q}^{*};U;\omega \right) =-\partial \Gamma ^{\left( 0,2\right) }_{Rzz\left( c\right) }\left( q_{0},\left| \mathbf{q}\right| =2k_{F}:U;\omega \right) /\partial \ln q_{0} \)
satisfies the RG equation

\begin{equation}
\label{34}
\left( \omega \frac{\partial }{\partial \omega }+\sum _{i}\beta _{i}\left( U_{x},U_{C},U_{f}\right) \frac{\partial }{\partial U_{i}}-2\gamma _{\chi _{s\left( c\right) }}\right) \Phi ^{\left( 0,2\right) }_{Rzz\left( c\right) }\left( q_{0};\mathbf{q}^{*};U;\omega \right) =0
\end{equation}
where \( \gamma _{\chi }=\omega d\ln Z_{\chi }\left( \mathbf{q}*;\omega \right) /d\omega  \).
Up to one-loop order as indicated in \( Fig.6 \) there is no mixing of channels.
Thus we assume that the zero-order coupling is associated with only one of the
existing scattering channels. It then follows that taking into account that
\( \Phi ^{\left( 0,2\right) }_{Rzz} \) is dimensionless the general solution
of this RG equation reduces to

\begin{equation}
\label{35}
\Phi ^{\left( 0,2\right) }_{Rzz\left( c\right) }\left( q_{0};\mathbf{q}^{*};\left\{ U_{i}\right\} ;\omega \right) =F_{s\left( c\right) }\exp \left( -2\int _{\omega }^{q_{0}}d\ln \left( \frac{\overline{q_{0}}}{\omega }\right) \gamma _{\chi _{s\left( c\right) }}\left[ \left\{ U_{i}\left( \overline{q_{0}};U_{i}\right) \right\} \right] \right) 
\end{equation}
with

\begin{equation}
\label{36}
\frac{dU_{i}\left( \overline{q_{0}},\left\{ U_{i}\right\} \right) }{d\ln \left( \frac{\overline{q_{0}}}{\omega }\right) }=\beta _{i}\left( U_{i}\left( \overline{q_{0}};\left\{ U_{i}\right\} \right) \right) 
\end{equation}
and  \( U_{i}\left( \overline{q_{0}}=\omega ;\left\{ U_{i}\right\} \right) =U_{i}\left( \omega \right)  \).
To determine both \( F_{s\left( c\right) } \) and \( \gamma _{\chi _{s\left( c\right) }} \)
we have to invoke perturbation theory. We can do this using the perturbation
expansion for the bare function \( \Gamma ^{\left( 2,1\right) }_{0z\left( c\right) } \)
together with the appropriate \( RG \) condition for the renormalized \( \Gamma ^{\left( 2,1\right) }_{R} \).
Since \( \Gamma ^{\left( 2,1\right) }_{0\uparrow \uparrow }\left( x,y;z\right) =G_{0\uparrow }^{-1}\left( x-z\right) G_{0\uparrow }^{-1}\left( z-y\right) G^{\left( 2,1\right) }_{0\uparrow \uparrow z\left( c\right) }\left( x,y;z\right)  \)
where \( G^{\left( 2,1\right) }_{0\uparrow \uparrow z\left( c\right) }\left( x,y;z\right)  \)
is defined as

\begin{equation}
\label{37}
G^{\left( 2,1\right) }_{0\uparrow \uparrow z\left( c\right) }\left( x,y;z\right) =\left\langle \psi _{0\uparrow }\left( x\right) \psi _{0\uparrow }^{\dagger }\left( y\right) \xi _{s\left( c\right) }\left( z\right) \exp -\int _{w}\mathcal{L}_{int}\left[ \psi _{0\sigma }^{\dagger }\left( w\right) ,\psi _{0\sigma }\left( w\right) ;U_{0}\right] \right\rangle 
\end{equation}
it follows that \( \left( Fig.7\right)  \)

\begin{equation}
\label{38}
\Gamma ^{\left( 2,1\right) }_{0\uparrow \uparrow z\left( c\right) }\left( q\right) =1\mp U_{0}\int _{p}G_{0\downarrow }\left( p+q\right) G_{0\downarrow }\left( p\right) +...
\end{equation}
Using \( RG \) theory the bare \( \Gamma ^{\left( 2,1\right) }_{0\uparrow \uparrow z\left( c\right) } \)
is related to the corresponding renormalized \( \Gamma ^{\left( 2,1\right) }_{R\uparrow \uparrow z\left( c\right) } \)
by

\begin{equation}
\label{39}
\Gamma ^{\left( 2,1\right) }_{0\uparrow \uparrow z\left( c\right) }\left( p;q\right) =Z_{s\left( c\right) }\left( \mathbf{q}^{*};\omega \right) Z^{-\frac{1}{2}}\left( \mathbf{p}+\mathbf{q};\omega \right) Z\left( \mathbf{p};\omega \right) ^{-\frac{1}{2}}\Gamma ^{\left( 2,1\right) }_{R\uparrow \uparrow z\left( c\right) }\left( p;q;\omega \right) 
\end{equation}
 If we define the renormalized \( \Gamma ^{\left( 2,1\right) }_{R\uparrow \uparrow z\left( c\right) } \)
such that \( \Gamma ^{\left( 2,1\right) }_{R\uparrow \uparrow z\left( c\right) }\left( q_{0}=\omega ,\mathbf{q}=\mathbf{q}^{*};U\right) =1 \)
and taking into consideration that to order \( O\left( U_{0}\right)  \) , \( Z\left( \mathbf{p}+\mathbf{q};\omega \right) =Z\left( \mathbf{p};\omega \right) =1 \)
it follows immediately that

\begin{equation}
\label{37}
Z_{\chi _{s\left( c\right) }}\left( \mathbf{q}^{*};\omega \right) =1\mp \frac{\lambda -\Delta -\delta }{2\pi ^{2}k_{F}}U_{0}\ln \left( \frac{\Omega }{\omega }\right) +...,
\end{equation}
and \( F_{s}=F_{c}=\left( \lambda -\Delta -\delta \right) /\pi ^{2}k_{F} \).
As we can see from \( Fig.7 \) the coupling function involved in this process
is such that it does not allow the exchange of spins between left and right
ingoing or outgoing legs. Following our regularization scheme we can then relate
\( U_{0} \) to an appropriate renormalized forward coupling function \( U_{f}\left( \omega \right)  \).
Assuming that as the scale parameter \( \omega \rightarrow 0 \) this running
coupling function approaches a non-zero fixed point \( U_{f}\left( \omega \right) \rightarrow U_{f}^{*} \)
, \( \gamma _{\chi } \) becomes

\begin{equation}
\label{38}
\gamma _{\chi _{s\left( c\right) }}=\pm \frac{\lambda -\Delta -\delta }{2\pi ^{2}k_{F}}U_{f}^{*}\pm ...
\end{equation}
Hence, in the vicinity of the Fermi Surface , for \( \mathbf{q}=\mathbf{q}^{*} \),
\( \Phi ^{\left( 0,2\right) }_{Rzz} \) reduces to

\begin{equation}
\label{39}
\Phi ^{\left( 0,2\right) }_{Rzz\left( c\right) }\left( q_{0};\mathbf{q}^{*};U_{f}^{*};\omega \right) =\frac{\lambda -\Delta -\delta }{\pi ^{2}k_{F}}\left( \frac{q_{0}}{\omega }\right) ^{\mp \left( \frac{\lambda -\Delta -\delta }{\pi ^{2}k_{F}}\right) U_{f}^{*}}
\end{equation}
If we integrate this out with respect to \( \ln \left( \frac{q_{0}}{\omega }\right)  \)
we find that the corresponding renormalized spin( charge )susceptibility \( \Gamma ^{\left( 0,2\right) }_{Rzz\left( c\right) } \)
is given in the form

\begin{eqnarray}
\chi _{s\left( c\right) }\left( q_{0};\mathbf{q}^{*};\omega \right) =\Gamma ^{\left( 0,2\right) }_{Rzz\left( c\right) }\left( q_{0};\mathbf{q}^{*};U_{f}^{*};\omega \right)  &  & \\
=\pm \frac{1}{U_{f}^{*}}\left[ \left( \frac{q_{0}}{\omega }\right) ^{\mp \left( \frac{\lambda -\Delta -\delta }{\pi ^{2}k_{F}}\right) U_{f}^{*}}-1\right]  &  & \label{40} 
\end{eqnarray}
Clearly \( \Gamma ^{\left( 0,2\right) }_{Rzz} \) is singular if \( U_{f}^{*}<0 \)
for \( q_{0}>0 \) and \( \omega \rightarrow 0 \). In contrast, in this limit,
\( \Gamma ^{\left( 0,2\right) }_{Rc} \) reduces to \( 1/\left| U^{*}_{f}\right|  \).
The renormalized charge susceptibility is therefore finite for \( q_{0}>\omega \rightarrow 0 \).
However if we consider the mixing of channels effect in higher order perturbation
theory which nullifies the fixed forward coupling \( U^{*}_{f} \) the spin
and charge susceptibilities in one-loop order reduce to

\begin{equation}
\label{41}
\chi _{s\left( c\right) }\left( q_{0};\omega \right) =-\left( \frac{\lambda -\Delta -\delta }{\pi ^{2}k_{F}}\right) \ln \left( \frac{q_{0}}{\omega }\right) 
\end{equation}
 Nevertheless if we consider those higher order effects in the fixed couplings
for consistency we must also take into account the corresponding higher loop
contributions for both \( Z_{\chi _{s\left( c\right) }} \)and \( \Phi _{Rzz\left( c\right) }^{\left( 0,2\right) } \).
This brings important modifications to this limit and will be discussed elsewhere. 

Consider next the pairing susceptibility \( \chi _{p}\left( q\right) =\Gamma ^{\left( 0,2\right) }_{p}\left( q\right)  \)
for \( \left| \mathbf{q}\right| =0 \) defined by

\begin{equation}
\label{41}
\chi _{p}\left( q\right) =\Gamma ^{\left( 0,2\right) }_{p}\left( q\right) =\frac{\delta ^{2}}{\delta \eta \left( q\right) \delta \eta ^{\dagger }\left( q\right) }\left\langle \exp -\int _{k,p}\psi ^{\dagger }_{\uparrow }\left( p\right) \psi ^{\dagger }_{\downarrow }\left( -p+k\right) \eta \left( k\right) +h.c.\right\rangle 
\end{equation}

If we follow the same steps as before using the fact that \( \Pi ^{\left( 0\right) }\left( q_{0},\left| \mathbf{q}\right| =0\right) =\Pi ^{\left( 0\right) }\left( q_{0}\right)  \)
is logarithmic divergent in the 'Cooper' channell we define the renormalized
pairing correlation function \( \Gamma ^{\left( 0,2\right) }_{Rp}\left( q_{0};U;\omega \right)  \)
by

\begin{equation}
\label{42}
\Gamma ^{\left( 0,2\right) }_{Rp}\left( q_{0};\mathbf{q}=0;\left\{ U_{i}\right\} ;\omega \right) =Z_{\chi _{p}}^{2}\left( \mathbf{q}=0;\omega \right) \Gamma ^{\left( 0,2\right) }_{0p}\left( q_{0};\mathbf{q}=0;\left\{ U_{0i}\right\} \right) +B\left( \omega \right) ,
\end{equation}
where the corresponding bare correlation function, up to order \( O\left( U_{0C}\right)  \)
in perturbation theory is \( \left( Fig.8\right)  \)

\begin{equation}
\label{43}
\Gamma ^{\left( 0,2\right) }_{0p}\left( q_{0};\mathbf{q}=0;U_{0}\right) =\Pi ^{\left( 0\right) }\left( q_{0};\mathbf{q}=0\right) -U_{0C}\left( \Pi ^{\left( 0\right) }\left( q_{0};\mathbf{q}=0\right) \right) ^{2}+...
\end{equation}
Defining the corresponding \( \Phi ^{\left( 0,2\right) }_{Rp}\left( q_{0};\mathbf{q}=0;U_{C};\omega \right) =\frac{\partial }{\partial \ln q_{0}}\Gamma ^{\left( 0,2\right) }_{Rp}\left( q_{0};\mathbf{q}=0;U_{C};\omega \right)  \)
it satisfies a RG type equation similar to the corresponding \( RG \) equation
for \( \Phi ^{\left( 0,2\right) }_{Rzz\left( c\right) } \). Thus in one-loop
order we can write its formal solution as

\begin{equation}
\label{44}
\Phi ^{\left( 0,2\right) }_{Rp}\left( q_{0};\mathbf{q}=0;U_{C};\omega \right) =A\exp \left( -2\int d\ln \left( \frac{\overline{q}_{0}}{\omega }\right) \gamma _{\chi _{p}}\left[ U_{C}\left( \overline{q}_{0};U\right) \right] \right) ,
\end{equation}
with

\begin{equation}
\label{45}
\gamma _{\chi _{p}}=\omega \frac{d\ln Z_{\chi _{p}}\left( \omega \right) }{d\omega }
\end{equation}
Once again we can use perturbation theory to determine both \( A \) and \( Z_{\chi _{p}} \).
Defining as before \( \Gamma ^{\left( 2,1\right) }_{0p}\left( x,y;z\right) =-G_{0\uparrow }^{-1}\left( x-z\right) G_{0\downarrow }^{-1}\left( y-z\right) G^{\left( 2,1\right) }_{0p}\left( x,y;z\right)  \)
with \( G^{\left( 2,1\right) }_{0p}\left( x,y;z\right)  \) given by

\begin{equation}
\label{45}
G^{\left( 2,1\right) }_{0p}\left( x,y;z\right) =\left\langle \psi _{0\uparrow }\left( x\right) \psi _{0\downarrow }\left( y\right) \psi _{0\uparrow }^{\dagger }\left( z\right) \psi ^{\dagger }_{0\downarrow }\left( z\right) \exp -\int \mathcal{L}_{int}\left[ \psi _{0\sigma },\psi ^{\dagger }_{0\sigma }\right] \right\rangle 
\end{equation}
we find that \( \left( Fig.9\right)  \)

\begin{eqnarray}
\Gamma ^{\left( 2,1\right) }_{0p}\left( q;U_{0C}\right) =1-U_{0C}\int _{p}G_{0\uparrow }\left( p\right) G_{0\downarrow }\left( -p+q\right) +... &  & \\
=Z_{\chi _{p}}\left( \mathbf{q}=0;\omega \right) Z^{-\frac{1}{2}}\left( -\mathbf{p};\omega \right) Z\left( \mathbf{p};\omega \right) ^{-\frac{1}{2}} & \Gamma ^{\left( 2,1\right) }_{Rp}\left( q;U_{C};\omega \right)  & \label{46} 
\end{eqnarray}
If we define the renormalized \( \Gamma ^{\left( 2,1\right) }_{Rp} \) such
that \( \Gamma ^{\left( 2,1\right) }_{Rp}\left( q_{0}=\omega ,\mathbf{q}=0;U_{C};\omega \right) =1 \)
it then follows that

\begin{equation}
\label{46}
Z_{\chi _{p}}\left( \omega \right) =1+\frac{4\lambda }{2\pi ^{2}k_{F}}U_{C}\left( \mathbf{q}=0;\omega \right) \ln \left( \frac{\Omega }{\omega }\right) +...
\end{equation}
and

\begin{equation}
\label{47}
A=\frac{4\lambda }{\pi ^{2}k_{F}}
\end{equation}
Thus if the physical system approaches criticality as \( \omega \rightarrow 0 \),
\( U_{C}\left( \omega \right) \rightarrow U_{C}^{*} \), \( \gamma _{\chi _{p}} \)
reduces to

\begin{equation}
\label{48}
\gamma _{\chi _{p}}=-\frac{4\lambda }{2\pi ^{2}k_{F}}U_{C}^{*}
\end{equation}
and as a result

\begin{equation}
\label{49}
\Phi ^{\left( 0,2\right) }_{Rp}\left( q_{0};\mathbf{q}=0;U^{*}_{C};\omega \right) =\frac{4\lambda }{\pi ^{2}k_{F}}\left( \frac{q_{0}}{\omega }\right) ^{\frac{4\lambda }{\pi ^{2}k_{F}}U_{C}^{*}}
\end{equation}
Finally integrating this result with respect to \( \ln \left( \frac{q_{0}}{\omega }\right)  \)
we find

\begin{eqnarray}
\chi _{p}\left( q_{0};\mathbf{q}=0;\omega \right) =\Gamma ^{\left( 0,2\right) }_{Rp}\left( q_{0};\mathbf{q}=0;U^{*};\omega \right)  &  & \\
=\frac{1}{U_{C}^{*}}\left[ \left( \frac{q_{0}}{\omega }\right) ^{\frac{4\lambda }{\pi ^{2}k_{F}}U_{C}^{*}}-1\right]  &  & \label{50} 
\end{eqnarray}
It is clear that \( \Gamma ^{\left( 0,2\right) }_{Rp} \) is finite and reduces
to \( -1/U_{C}^{*} \) when \( \omega \rightarrow 0 \) and \( q_{0}>0 \).
The singularity for \( q_{0}\geq \omega \rightarrow 0 \) in the bare pairing
susceptibility \( \Gamma ^{\left( 0,2\right) }_{0p} \) is therefore cancelled
out exactly by the renormalization factor \( Z_{\chi _{p}}\left( \omega \right)  \).
The non-Fermi liquid phase at two-loop level is therefore further characterized
by finite charge and pairing susceptibilities and a singular spin susceptibility
which might well represent the existence of a charge pseudogap and the absence
of a superconducting regime together with strong spin fluctuations. 

Notice however that all this changes as we move around in \textbf{\( \mathbf{k}- \)}space
taking into consideration even approximately the effects produced by the mixing
of scattering channels. To see this in a brief schematic form we can define
the renormalized external parameter \( \theta  \) which drives the physical
system towards the critical regime such that

\begin{equation}
\label{52}
\theta =Z_{\chi _{p}}^{-1}\left( \theta _{0}-\theta _{0c}\right) 
\end{equation}
where \( \theta _{0} \) and \( \theta _{0c} \) are the corresponding bare
terms which appears in the renormalized lagrangian as we define a source term
in terms of the Cooper pairing susceptibility. It follows from this that \( \theta  \)
satisfies the \( RG \) equation

\begin{equation}
\label{53}
q_{0}\frac{d\theta \left( q_{0},\theta \right) }{dq_{0}}=-\gamma _{\chi _{p}}\left[ U_{C}\left( q_{0},U_{C}\right) \right] \theta \left( q_{0},\theta \right) 
\end{equation}
with \( \theta \left( q_{0}=\omega ,\theta \right) =\theta  \) and \( U_{C}\left( q_{0}=\omega ,U_{C}\right) =U_{C} \).
Integrating this out we get immediately

\begin{equation}
\label{54}
\theta \left( q_{0},\theta \right) =\theta \exp -\int ^{q_{0}}_{\omega }\gamma _{\chi _{p}}\left[ U_{C}\left( \overline{q}_{0},U_{C}\right) \right] \frac{d\overline{q}_{0}}{\overline{q}_{0}}
\end{equation}
Choosing the \( q_{0} \) such that \( \theta \left( q_{0},\theta \right) =q_{0} \)
we obtain the scaling relation

\begin{equation}
\label{55}
\frac{\theta }{\omega }=\exp \int ^{q_{0}}_{\omega }(1+\gamma _{\chi _{p}}\left[ U_{C}\left( \overline{q}_{0},U_{C}\right) \right] )\frac{d\overline{q}_{0}}{\overline{q}_{0}}
\end{equation}
Since as we approach the critical region \( \gamma _{\chi _{p}}\left[ U_{C}\left( \overline{q}_{0},U_{C}\right) \right] \rightarrow \gamma _{\chi _{p}}\left( U_{C}^{*}\right)  \)
this reduces to

\begin{equation}
\label{56}
\frac{\theta }{q_{0}}=\left( \frac{q_{0}}{\omega }\right) ^{\gamma _{\chi _{p}}\left( U^{*}_{C}\right) }
\end{equation}
However using our previous results and the fact the the coupling function are
dimensionless we must have that \( q_{0}\sim \theta ^{^{\frac{3}{8}\zeta }}=\theta ^{\nu } \)
with \( \nu =\frac{3}{8}\zeta <1 \). Thus it follows from this that

\begin{equation}
\label{57}
\left( \frac{q_{0}}{\omega }\right) ^{-2\gamma _{\chi _{p}}\left( U_{C}^{*}\right) }\sim \theta ^{-2\left( 1-\nu \right) }
\end{equation}
and therefore both \( \Phi ^{\left( 0,2\right) }_{Rp} \) and \( \chi _{p}\rightarrow \infty  \)
as a power law when \( \theta \rightarrow 0 \) signalling the superconducting
transition when \( \theta  \) is tunned to drive the physical system towards
criticality.

\section{Conclusion}

We present a two-loop field-theoretical renormalization group calculation of
a two-dimensional truncated Fermi Surface. Our Fermi Surface model consists
of four disconnected patches with both flat pieces and conventionally curved
arcs centered around \( \left( 0,\pm k_{F}\right)  \) and \( \left( \pm k_{F},0\right)  \)
in \( \mathbf{k} \)-space. Two-dimensional Fermi liquid like states are defined
around the central region of each patch. In contrast the patch border regions
are flat and as a result their associated single particle states have linear
dispersion law. These flat sectors are introduced specifically to produce nesting
effects which in turn generate logarithmic singularities in the particle-hole
channels that give non-Fermi liquid effects. In this way conventional 2\( d \)
Fermi liquid states are sandwiched by single particles with a linear dispersion
law to simulate the so-called ``cold'' spots as in the experimentally observed
truncated Fermi Surface of the underdoped normal phase of the high-temperature
superconductors. Our main motivation here is to test to what extent Fermi liquid
theory is applicable in the presence of flat Fermi surface sectors which are
indicative of a strong coupling regime. New experimental data on both optimally
doped and underdoped Bi2212\cite{Valla} above \( T_{c} \) indicate that the
imaginary part of the self-energy \( Im\Sigma \left( \omega \right)  \) scales
linearly with \( \omega  \) even along the \( \left( 0,0\right)  \)-\( \left( \pi ,\pi \right)  \)direction.
This is consistent with other photoemission experiments\cite{Valla} which support
a marginal Fermi liquid phenomenology over the whole Fermi Surface. Our results
are in general agreement with those experimental findings since the power law
corrections we find for this linear behavior can in some cases be so small as
not to be detectable by the present day experiments. Using perturbation theory
we calculate the two-loop self-energy of a single particle associated with a
curved \( FS \) sector. We find that the bare self-energy is such that \( Im\Sigma _{o}\left( \omega \right) \sim \omega  \)
and as a result \( Re\Sigma _{0}\left( \omega \right) \sim \omega \ln \left( \frac{\Omega }{\omega }\right)  \)
for \( \omega \sim 0 \) reproducing the marginal Fermi liquid phenomenology
at \( FS \). We calculate \( \Sigma _{0} \) as a function of both frequency
and momentum. It turns out that both \( \partial \Sigma _{0}/\partial p_{0} \)
and \( \partial \Sigma _{0}/\partial \overline{p} \) diverge at \( FS \).
Using \( RG \) theory we determine the renormalized one-particle irreducible
function \( \Gamma ^{\left( 2\right) }_{R}\left( p_{0},\mathbf{p};U,\omega \right)  \)
in the vicinity of a ``cold'' spot \( FS \) point. Again it follows immediately
that the quasiparticle weight \( Z \) vanishes identically at the Fermi Surface.
Next we calculate the bare one-particle irreducible two-particle function \( \Gamma ^{\left( 4\right) }_{0\alpha \beta }\left( p_{1},p_{2};p_{3},p_{4}\right)  \)
for \( \alpha ,\beta =\uparrow ,\downarrow  \). This function depends on the
spin arrangements as well as on the relative momenta of its external legs. There
are three different scattering channels associated with the \( \Gamma ^{\left( 4\right) }_{0\alpha \beta } \)'s
: the so-called Cooper, exchange and forward channels. For a \( FS \) with
flat sectors there are logarithmic singularities in \( \Gamma ^{\left( 4\right) }_{0\alpha \beta } \)
for both exchange and forward channels due to nesting effects. In contrast the
Cooper channel produces similar singularities in the whole \( FS \). We calculate
\( \Gamma ^{\left( 4\right) }_{0} \) perturbatively up to two-loop order for
the mentioned scattering channels. Taking into account self-energy corrections
calculated earlier on we obtain the corresponding renormalized one-particle
irreducible function \( \Gamma ^{\left( 4\right) }_{R} \) subjected to an appropriate
renormalization condition. The field theory regularization scheme allow us to
introduce local counterterms to cancels all divergences order by order in perturbation
theory. This simplifies the problem considerably although due to the anisotropy
in \( \mathbf{k}- \) space the counterterms are in fact momenta dependent.
The bare coupling constant becomes a bare coupling function and we proceed with
the regularization of the divergences grouping them together accordind to their
location at \( FS \), the scattering channel and the vertex type interaction.
We introduce in this way three bare coupling functions \( U_{0x}\left( \mathbf{p}\right)  \),
\( U_{0C}\left( \mathbf{p}\right)  \) and \( U_{0f}\left( \mathbf{p}\right)  \)
to systematically cancel all the local divergences in our perturbation theory
expansions. In this scheme the effects produced by the crossing of different
channels is practically non-existent up to the two-loop order calculation of
the self-energy. Therefore if we only consider the leading divergence at every
order of perturbation theory it is essential to proceed with the renormalization
of fields in order for the running coupling functions to develop non-trivial
infrared \( \left( IR\right)  \) stable fixed points. As opposed to the parquet
or the Wilsonian \( RG \) methods we don't try to derive an effective action
in explicit form. This would in practice demand the introduction of an infinite
number of local counterterms in our lasgrangian model. Nevertheless all divergences
can be removed to all orders by local subtractions around a given point of momentum
space. The anisotropy of the Fermi Surface therefore reflects itself directly
in the momentum dependence of the coupling parameters. This feature is consistent
with the findings of those other two approaches refered before. However since
we take explicit into account self-energy even if we don't go beyond the leading
divergence approximation and we are able to find non-trivial fixed point solutions
in our higher loop calculations. With our two-loop results for the non-trivial
fixed points together with the assumption that the physical system acquires
a critical condition as we approach the Fermi Surface by taking the scale parameter
\( \omega \rightarrow 0 \) we can solve the \( RG \) equation for the renormalized
single particle propagator \( G_{R}\left( p;U,\omega \right)  \) in the vicinity
of a chosen ``cold'' spot point. We show that the nullification of the quasiparticle
weight \( Z \) manifests itself as an anomalous dimension in \( G_{R} \).
This anomalous dimension is independent of the sign of the given fixed point
value. Using this result we calculate the spectral function \( A_{R}\left( p;U^{*},\omega \right)  \)
and the renormalized single particle dispersion law. From this we calculate
the ``Fermi velocity'' which can either remain finite or is nullified at \( FS \).
Finally we calculate the corresponding momentum distribution function \( n\left( \overline{p}\right)  \)
in the vicinity of \( FS \) and show that it is a continuous function of \( \overline{p} \)
with \( \partial n\left( \overline{p}\right) /\partial \overline{p} \) finite
or \( \partial n\left( \overline{p}\right) /\partial \overline{p}\rightarrow \infty  \)
when \( \overline{p}/\omega \rightarrow 0 \). In the former case we have a
real charge gap typical of an insulating state and in the latter the physical
system continues to be metallic and resembles a Luttinger liquid. There could
be in this way phase separation in \( \mathbf{k}- \) space and there is a complete
breakdown of the Landau Fermi liquid when the ``cold'' spot suffers the effects
produced by the flat \( FS \) sectors. Since essentially the \( RG \) exponentiates
the self-energy \( \ln  \) corrections the power law behavior of \( G_{R} \)
reflects itself back in the renormalized self-energy given

\begin{equation}
\label{130}
Im\Sigma _{R}\left( \mathbf{p};U^{*}\right) \sim -\left| p_{0}\right| \left( \frac{p_{0}^{2}}{\omega ^{2}}\right) ^{-\frac{\gamma ^{*}\left( \mathbf{p};U^{*}\right) }{2}}
\end{equation}
with \( \gamma ^{*} \)depending on the size of the size of the flat Fermi Surface
sectors through the fixed point coupling strength. For certain \( \mathbf{k}- \)space
regions near \( FS \) we show that \( \gamma ^{*}\simeq 6/121 \). This produces
a singular \( \partial n\left( \overline{p}\right) /\partial \overline{p}\rightarrow \infty  \)
and is consistent with the marginal Fermi liquid phenomenology which is in agreement
with the observed experimental results.

Since several of our results are given in terms of a fixed point value it is
important to see what happens if we consider higher order contributions to our
perturbation theory expansions. We do this calculating initially the 3-loop
corrections to the bare self-energy \( \Sigma _{0\uparrow } \). At this order
of perturbation theory we have the bare constants \( U_{0C}^{3}\cong U_{0x}^{3}\cong U_{0f}^{3} \)
identical to each other. As a result the two existing contributions in this
order cancel each other exactly for \( p_{0}\sim \omega \sim 0 \). However
if we consider 3-loop terms in our perturbative calculation of the three bare
couplings we can distinguish the different contributions produced at order \( O\left( U^{2}_{0}\right)  \)
and this brings important changes to our results. The exchange and Cooper channel
couplings mix strongly with each other. As a result of this we define generalized
\( \beta  \)-functions by considering \( \beta _{i}=\omega \partial U_{i}/\partial \omega =\beta _{i}\left( U_{C},U_{x},U_{f}\right)  \)
for \( i=C,x,f \). We calculate the eigenvalue matrix \( M_{ij}=\frac{\partial \beta _{i}}{\partial U^{*}_{j}} \),
find its eigenvalues and expand these \( \beta  \)-functions in coupling space
around one of the existing fixed points in \( \mathbf{k}- \)space. We then
show that the critical surface formed by the set of trajectories of \( U_{i}\left( \mathbf{p};\omega \right)  \)
which is attracted into the fixed point \( \left( U_{C}^{*}\left( \mathbf{p}\right) ,U_{x}^{*}\left( \mathbf{p}\right) ,U_{f}^{*}\left( \mathbf{p}\right) \right)  \)
for \( \omega \rightarrow 0 \) has codimensionality one. This means that one
external parameter such as temperature or hole concentration is needed to drive
the physical system towards its phase transition characterized by its fixed
points \( \left( U_{C}^{*},U_{x}^{*},U_{f}^{*}\right)  \)'s. 

To enquire about the nature of our non-Fermi state and to find out how it reacts
to its various instabilities we calculate the longitudinal susceptibility \( \chi _{s} \)
, the charge susceptibility \( \chi _{c} \) and the pairing susceptibility
\( \chi _{p} \) around a given point of momentum space. These functions are
obtained from appropriate \( RG \) equations which are solved perturbatively
if the renormalized running coupling functions are precisely at the non-trivial
fixed point. For simplicity we use perturbation theory up to one-loop order
and as a result there is no strong mixing of scattering channels. Both \( \chi _{s} \)
and \( \chi _{c} \) are linked to the forward scattering channel. Since \( U_{f}^{*}<0 \)
in two-loop order the corresponding multiplicative renormalization constants
\( Z_{\chi _{s}} \) and \( Z_{\chi _{c}} \) produce cancellation effects in
\( \chi _{c}\left( q_{0};\mathbf{q};\omega \right)  \) but not in \( \chi _{s}\left( q_{0};\mathbf{q};\omega \right)  \)
since it continues to have a power law singularity for \( q_{0}>\omega  \)
and \( \omega \rightarrow 0 \). Moreover in this order of perturbation theory
we show that the pairing susceptibility \( \chi _{p}\left( q_{0};\mathbf{q};\omega \right)  \)
is finite at the fixed point for \( q_{0}>\omega  \) and \( \omega \rightarrow 0 \)
since \( U^{*}_{C} \)is non-zero and the Cooper singularity is cancelled exactly
by the renormalization factor \( Z_{\chi _{p}} \). This result depends on the
location of the interacting particles at the Fermi Surface and as a result it
changes as we move around in \( \mathbf{k}- \) space reflecting strong anisotropy
effects. Nevertheless this shows that this corresponding non-Fermi liquid state
is non-superconducting with a finite \( \chi _{c} \) which could be indicative
of charge incompressibility or charge pseudo gap behavior \cite{Zanchi2} together
with strong spin fluctuations. As a matter of fact an external parameter \( \theta  \)
must be tuned for the system to approach the critical region and the superconducting
transition. Using a simple scaling argument we then show that in this case the
pairing susceptibility \( \chi _{p} \) diverges as a power law when \( \theta \rightarrow 0 \)
in the superconducting critical point.

To conclude it is fair to say that even a simplified anisotropic Fermi Surface
model as the one used in this work is able to produce interesting non-trivial
results due to the fact that it contains flat parts which are indicative of
a strong interaction regime. They can turn \( 2d \)-Fermi liquid states into
a non-Fermi liquid which can be metallic or insulating depending on its location
at the fermi Surface in \( \mathbf{k}- \)space. It is tempting to relate our
results to the high \( T_{c} \) superconductors which are known to have an
anisotropic \( FS \) with a pseudogap which show non-Fermi liquid behavior
for the underdoped and optimally doped metallic phase above the critical temperature.
Our findings concerning the nature of the metallic state are in general agreement
with more recent photoemission experiments which demonstrate the validity of
the marginal Fermi liquid phenomenology above \( T_{c} \). We believe therefore
the model presented here might well contain some of the ingredients which are
needed to describe the strange metal and the pseudogap phases of the cuprate
superconductors.

\emph{Acknowledgements-} I am grateful to T.M.Rice, C.Honerkamp and K.Voelker
\( \left( ETH-Zurich\right)  \) and D.Baeriswyl, B.Binz, C.Morais-Smith and
C.Aebischer \( \left( ITP-Fribourg\right)  \) for several stimulating discussions.
Discussions with Peter Kopietz and with Benham Farid are also greatly appreciated.
Finally I also want to thank the Center for Theoretical Studies-ETH-Zurich,
the Institute de Physique Theorique-Univ. de Fribourg and the Swiss National
Foundation for their very nice hospitality. This work is partially supported
by the CNPq \( \left( Brazil\right)  \).

{*} \emph{permanent address:} Centro Intl. de Fis. da Mat. Condensada-UnB, Brasilia-DF
\( \left( Brazil\right)  \).

Figure Captions:

Fig. 1- Truncated Fermi Surface Model

Figs. 2- Feynmann diagrams up to 2-loop order for \( \Gamma ^{\left( 4\right) }_{\uparrow \downarrow } \)in
the: \( \left( a\right)  \) exchange channel, \( \left( b\right)  \)Cooper
channel and \( \left( c\right)  \) forward channel

Fig.3- Diagrams up two-loop order for \( \Gamma ^{\left( 4\right) }_{\uparrow \uparrow } \)

Fig.4- Diagrams for \( \Sigma _{\uparrow } \) in 3-loop order

Fig.5- 4-loop diagrams for \( \Sigma _{\uparrow } \)

Fig.6-\( \Gamma ^{\left( 0,2\right) }_{0zz\left( c\right) } \) up to one-loop
order

Fig.7-\( \Gamma ^{\left( 2,1\right) }_{0zz\left( c\right) } \) up to one-loop
order

Fig.8-\( \Gamma ^{\left( 0,2\right) }_{0p} \) up to one-loop order

Fig.9- \( \Gamma ^{\left( 2,1\right) }_{0p} \) up to one-loop order

\end{document}